\documentclass[%preprint,
superscriptaddress,
 amsmath,amssymb,
 aps,
reprint
]{revtex4-1}

\usepackage{color}
\usepackage{graphicx}% Include figure files
\usepackage{dcolumn}% Align table columns on decimal point
\usepackage{bm}% bold math
\usepackage{epstopdf}
\usepackage{gensymb}
\usepackage{hyperref}
\usepackage{float}

\begin{document}

%\preprint{APS/123-QED}

\title{Perspective: strain and strain gradient engineering in membranes of quantum materials}% Force line breaks with \\

\author{Dongxue Du}
\author{Jiamian Hu} \email{jhu238@wisc.edu}
\author{Jason K. Kawasaki} \email{jkawasaki@wisc.edu}
\affiliation{Materials Science and Engineering, University of Wisconsin-Madison, Madison, WI 53706}

\date{\today}% It is always \today, today,

\begin{abstract}
Strain is powerful for discovery and manipulation of new phases of matter; however, the elastic strains accessible to epitaxial films and bulk crystals are typically limited to small ($<2\%$), uniform, and often discrete values. This Perspective highlights new directions for strain and strain gradient engineering in free-standing single crystalline membranes of quantum materials. Membranes enable large ($\sim 10\%$), continuously tunable strains and strain gradients via bending and rippling. Moreover, strain gradients break inversion symmetry to activate polar distortions, ferroelectricity, chiral spin textures, novel superconductivity, and topological states. Recent advances in membrane synthesis by remote epitaxy and sacrificial etch layers enable extreme strains in new materials, including transition metal oxides and Heusler compounds, compared to natively van der Waals (vdW) materials like graphene. We highlight new opportunities and challenges for strain and strain gradient engineering in membranes of non-vdW materials.

\end{abstract}

\maketitle

\section{Introduction}

The properties of quantum materials with highly localized $d$ and $f$ orbitals can be highly sensitive to changes in bond lengths, bond angles, local coordination, and symmetry. Strain is a powerful knob for tuning these parameters, with striking examples including strain-induced superconductivity in epitaxial RuO$_2$ films \cite{ruf2021strain}, strain-induced ferroelecticity in SrTiO$_3$ \cite{schlom2007strain}, and strain-induced changes in magnetic ordering in magnetic shape memory alloys \cite{song2020strain}. However, the strains accessible to bulk materials and epitaxial films are typically limited to $<2 \%$ before relaxation via dislocations \cite{matthews1974defects, people1985calculation,van1963crystal}. Moreover, in epitaxial films the strain is static and discrete, based on the lattice mismatch between particular film and substrate combinations. As a result, many quantum properties remain out of reach.

\begin{figure*}[ht]
    \centering
    \includegraphics[width=.8\textwidth]{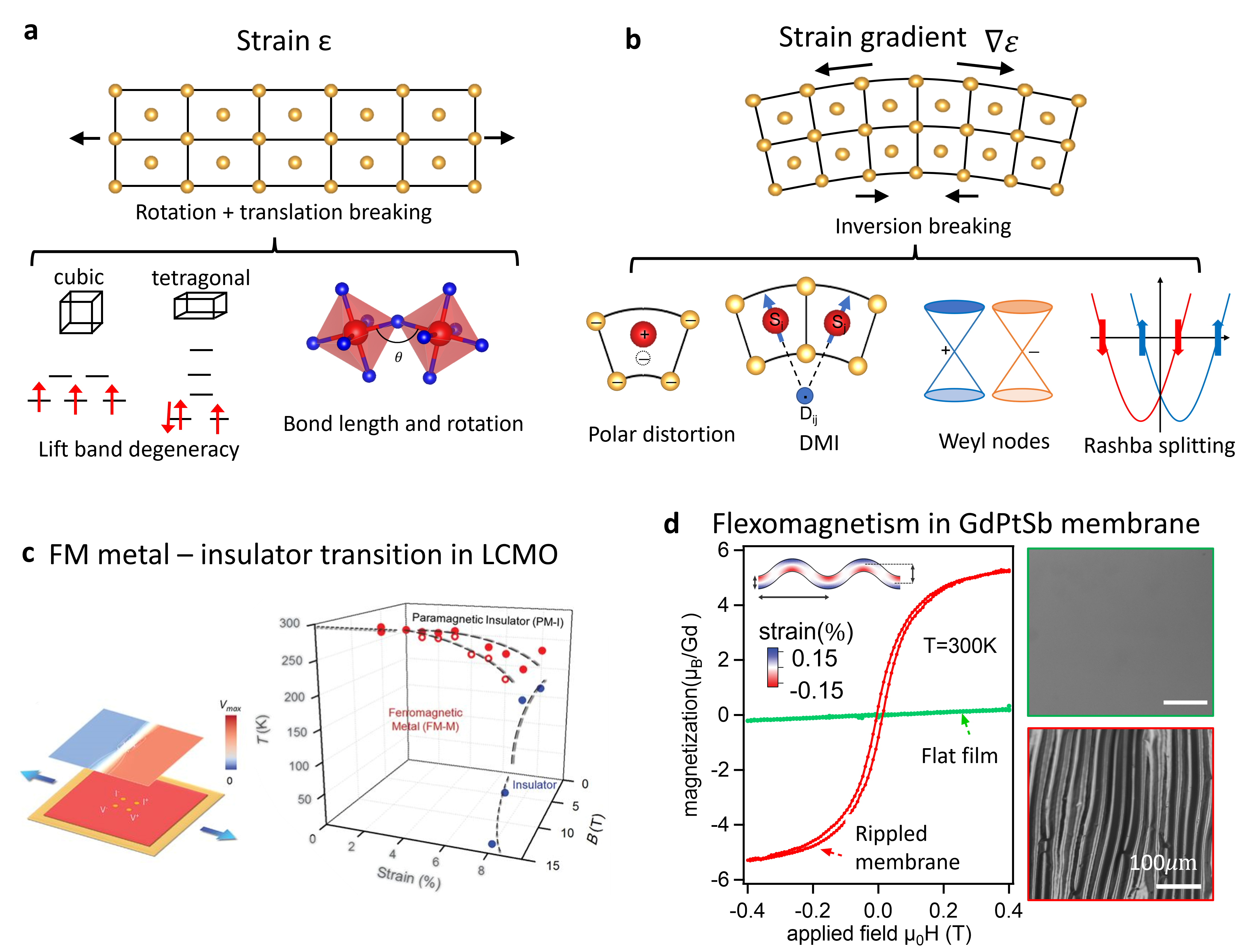}
    \caption{Symmetry breaking and properties induced by large strains and strain gradients. (a) Homogeneous strain breaks rotational and translational symmetry, to lift band degeneracies and tune bond lengths and angles. These parameters can tune magnetic exchange and electron correlations. (b) Strain gradients break inversion, providing access to polar distortions, tunable Dzaloshinskii-Moriya interaction (DMI), tunable Weyl nodes, and Rashba splitting. (c) Extremely large strain induced Metal to Insulator transition and magnetic phase transition in La$_{0.7}$Ca$_{0.3}$MnO$3$ membrane. From Hong, et. al. Science 368, 71 (2020) (Ref \cite{hong2020extreme}). Reprinted with permission from AAAS. (d) Flexomagnetism induced by strain gradients in rippled GdPtSb membranes. Reproduced from Du et. al. Nature Communications, 12, 2494 (2021) (Ref \cite{du2021epitaxy}), under Creative Commons licence.
    }
    \label{overview}
\end{figure*}

This Perspective highlights new opportunities for strain and strain gradient engineering in single crystalline membranes of quantum materials, beyond natively vdW materials. Free-standing membranes enable two regimes that are inaccessible in films and bulk crystals (Fig. \ref{overview}). First, membranes and other free-standing nanostructures can sustain much larger elastic strains (8\% in (La,Ca)MnO$_3$ membranes \cite{hong2020extreme} and 10\% in BaTiO$_3$ \cite{dong2019super}), compared to the $\sim 2\%$ limit for films and bulk crystals. Second, membranes enable controlled strain gradients via bending and rippling \cite{du2021epitaxy}. Whereas uniform strain breaks rotational and translational symmetries (Fig. \ref{overview}a), strain gradients break inversion symmetry. Inversion breaking is the necessary ingredient for ferroelectric polar distortions, nonlinear optical responses, Dzyaloshinskii-Moriya interaction (DMI)-induced chiral spin textures, and Rashba splitting (Fig. \ref{overview}b). 

Recent advances in remote epitaxy \cite{kim2017remote, kum2020heterogeneous, du2021epitaxy, du2022controlling, yoon2022free} and etch release layers \cite{hong2020extreme, dai2022highly} enable the synthesis of ultrathin membranes of quantum materials, including Heusler compounds and transition metal oxides. These synthesis advancements enable extreme strain to be applied to new classes of ultrathin membranes, which until recently were mainly limited to easily exfoliatable van der Waals (vdW) materials like graphene and transition metal dichalcogenides \cite{si2016strain,peng2020strain}. We highlight opportunities for discovery of new properties via large strains and strain gradients in these materials (Section II). Realization of these properties relies on new approaches for single crystalline membrane synthesis, understanding and controlling their extreme mechanical properties, and feedback from computational modelling (Section III). We conclude with an outlook on static and nonequilibrium stains (Section IV).

\section{Opportunities}

\subsection{Magnetism, flexomagnetism, and skyrmions} 
Homogeneous strains couple strongly to magnetism via piezomagnetism ($M \propto \epsilon$) and magnetostriction ($M^2 \propto \epsilon$). Microscopically, strain tunes magnetic exchange via the bond lengths and bond angles, and tunes the band degeneracies and occupancies via change in symmetry \cite{zhang2018strain}. Many Heusler compounds and transition metal oxides have rich magnetic properties \cite{casper2012half, bibes2007oxide, bibes2011ultrathin}. Larger strain in membranes have the potential to tune magnetic properties more substantially. As an example, $>5\%$ strains in (La,Ca)MnO$_3$ membranes induce a ferromagnetic metal metal to insulator transition \cite{hong2020extreme} (Fig. \ref{overview}c).

Magnetism can also couple strongly to strain gradients, which is termed flexomagnetism ($M \propto \nabla\epsilon$) \cite{eliseev2011linear, lukashev2010flexomagnetic, eliseev2009spontaneous}. Whereas strain gradients are difficult to control in bulk crystals and epitaxially clamped films, we recently demonstrated an antiferromagnetic to ferro/ferrimagnetic transition upon rippling in GdPtSb membranes, in the first experimental example of flexomagnetism \cite{du2021epitaxy} (Fig. \ref{overview}d). Although the microscopic mechanism is not well understood, we speculate that strain gradients enhance the DMI, leading to canted ferrimagnetism in the rippled GdPtSb membranes. Microscopic measurements and theory are required to understand the flexomagnetic response.

Inversion-breaking strain gradients can also tune or induce chiral spin textures such as skyrmions, via tuning the DMI. As proof of concept, recent experiments on partially relaxed (La,Sr)MnO$_3$ (LSMO) films show signatures of skyrmions, induced by inversion-breaking strain gradients along the growth direction \cite{zhang2021strain}. We anticipate even greater control of skyrmions in bent LSMO membranes, which allow the strain gradient to be tuned more precisely and continuously rather spontatneous strain relaxation in LSMO films.

Theory predicts skyrmions in other bent systems. Mesoscale calculations predict highly tunable skyrmions in bent membranes heterostructures of simple metals and a flexo-Hall effect induced by bending \cite{liu2022flexoresponses}.  More complex cyclical states are predicted for curved nanotubes of CrI$_3$, due to periodic boundary conditions along the circumference of the nanotube \cite{edstrom2022curved}. Strain gradients are also predicted to control skyrmion motion \cite{yanes2019skyrmion,liu2019manipulating,gorshkov2022dmi}, which could be controlled dynamically on a flexible membrane platform. We anticipate these concepts to apply broadly to membranes of new vdW materials, e.g. rare earth Heusler compounds, magnetic oxides, and chiral intermetallics.

\subsection{Ferroelectricity, flexoelectricity, and polar metals}

Ferroelectricity requires crystals with broken inversion symmetry that have a unique polar axis. Although homogeneous strain alone does not break inversion, it can tune ferroelecticity in systems that are already ferroelectric or induce ferroelectricity in materials on the verge of being ferroelectric. For example, uniaxial tensile strain induces ferroelectricity in membranes of the quantum paraelectric SrTiO$_3$, by suppressing quantum fluctuations \cite{xu2020strain}. The absence of strain (clamping) in membranes can also be important: free-standing BaTiO$_3$ membranes display faster switching than epitaxial BaTiO$_3$ films, due to the release from substrate clamping effects \cite{pesquera2020beyond}. 

Strain gradients, which break inversion, are even more powerful because they can induce polar distortions in materials that were originally centrosymmetric. The general coupling between ferroelectricity and strain gradients is termed flexoelectricity \cite{kogan1964piezoelectric, zubko2013flexoelectric, cross2006flexoelectric, indenbom1981flexoelectric}. Early experiments quantified the flexoelectric coefficients for few millimeter thick cantilevers of lead magnesium niobate \cite{ma2001observation} and lead zirconate titanate (PZT) \cite{ma2005flexoelectric}. More recent experiments suggest that 10 nm thick BaTiO$_3$ membranes released from graphene/Ge display an enhanced flexoelectric response compared to bulk \cite{dai2022highly}. We anticipate broader opportunities for flexoelectricity in ultrathin membranes, where the enhanced elasticity in the ultrathin limit provides access to new regimes for large strains and strain gradients.

Flexoelectric coupling may also enable the switching of polar metals. Unlike ferroelectic insulators, in which the electric polarization can be switched via an applied electric field, in polar metals the electric field is screened out by free carriers. Bending-induced strain gradients provides a means of switching a polar metal without application of an electric field \cite{zabalo2021switching}. First-principles calculations identified LiOsO$_3$ as a promising material for switching via strain gradients \cite{zabalo2021switching}. Other materials, including the high conductivity polar metals LaAuGe and GdAuGe\cite{du2019high, du2022controlling}, may also be good candidates.

Finally, ultrathin ferroelectric membranes provide opportunities for mechanically active materials, due to their extreme superelastic responses, large strains, and 180 degree bending. For BiFeO$_3$ membranes, 180 degree bending with 1 micron radius of curvature and reversible elastic strains up to 5.4\% are accommodated by a reversible rhombahedral-tetragaonal phase transformations \cite{peng2020phase}. For BaTiO$_3$ membranes, large bending strains of 10 percent are reported, which are enabled by continuous dipole rotations of ferroelectric domains \cite{dong2019super}. Similar arguments based on phase transformations and domain reorientations may apply for membranes of martensitic materials like shape memory alloys. These materials provide new opportunities for tuning stimuli-responsive materials that undergo large ferroelectric and ferroelastic phase transitions.

\subsection{Superconductivity.} Strain and strain gradients in membranes provide opportunities to enhance the critical temperature $T_c$ and critical fields of known superconductors, induce superconductivity in new materials, and tune the pairing symmetry and coupling to other electronic states such as topological states and ferroelectricity.

Strain can enhance the $T_c$ of known superconductors including iron based superconductors and cuprates \cite{engelmann2013strain}. For example, epitaxial strain enhances the upper critical field of Fe based superconductors \cite{tarantini2011significant}. In this family of materials, the strength of electronic correlations and $T_c$ are highly dependent on the X-Fe-X bond angle (X = pnictogen or chalcogen), with a maximum Tc when the bond angle is near 109 degrees \cite{okabe2010pressure, mandal2017correlated}. This bond angle is typically tuned by alloying, doping, or intercalation \cite{okabe2010pressure}, which introduces disorder. Freestanding membranes provide a path to cleanly and continuously tune the X-Fe-X bond angle via strain and bending-induced strain gradients. Decoupling a monolayer FeSe film from a SrTiO$_3$ substrate also enables the specific effects of interfacial-enhanced superconductivity to be tested \cite{wang2012interface, lee2014interfacial, tan2013interface}.

Strained membranes may provide similar opportunities for cuprates and other superconducting oxides. In cuprates, 0.5 \% compressive epitaxial strain nearly doubles the $T_c$ of (La,Sr)CuO$_4$ films from 25 K to 49 K \cite{locquet1998doubling}. One challenge for cuprates is that the Tc is also highly sensitive to oxygen stoichiometry \cite{bozovic2002epitaxial}, making it challenging to compare across separate samples. Membranes provide a possible solution for deconvolving stoichiometry from strain effects by allowing continuous tuning of strain on the same sample. Moreover, membranes enable larger strains and strain gradients.

The large strains and strain gradients in membranes also provide opportunities to induce superconductivity in new materials. Anisotropic strain induces superconductivity in RuO$_2$ films grown on TiO$_2$ (110) \cite{ruf2021strain}. Membranes provide further tunability for anisotropic strain, since the strain is not limited to particular film-substrate combinations and the strain in different crystallographic directions can be tuned independently. Inversion-breaking strain gradients may also an important new tool: for SrTiO$_3$, ferroelectric polar distortions are thought to stabilize superconductivity \cite{edge2015quantum}, and strain has been shown to tune the $T_c$ \cite{ahadi2019enhancing}. Tunable inversion breaking in membranes allows this idea to be tested in other classes of materials, beyond the few existing quantum paraelectrics \cite{muller1979srti, rytz1980dielectric}. 

Finally, inversion breaking strain gradients may find use for tuning the superconducting pairing symmetry \cite{yip2014noncentrosymmetric}. In noncentrosymmeric superconductors, mixtures of spin-singlet and spin-triplet pairing are allowed \cite{gor2001superconducting}, and interesting topological and magnetoelectric properties are expected \cite{yip2014noncentrosymmetric}. Membranes provide a means to continuously tune the crystalline symmetry of a superconducting material, to distinguish effects strain from disorder.

\subsection{Topological states.} 

Membranes provide an opportunity to tune topological band inversion and band gaps via large strains that are difficult to access in bulk materials or films. For example, FeSe$_{x}$Te$_{1-x}$ is suggested by ARPES and STM to be a topological superconductor for $x=0.45$ \cite{zhang2018observation, wang2018evidence, machida2019zero}, due to the strong spin-orbit coupling of Te and $p-d$ hybridization \cite{lohani2020band, wang2015topological}. Large strains in Fe(Se,Te) membranes may provide extended control of the band inversion and $p-d$ hybridization beyond what can be achieved by Te-alloying alone. As another example, whereas many rare earth half Heusler compounds are topological semimetals with overlapping valence and conduction bands \cite{lin2010half,chadov2010tunable,nakajima2015topological,singh2020rare}, it would be more attractive to have a material with a bulk bandgap. DFT calculations for LaPtBi suggest that a very large strain of $7\%$ is required to open a bulk band between overlapping $\Gamma_8-\Gamma_6$ band while preserving the band inversion \cite{al2010topological,xiao2010half}. This magnitude of strain is not possible in epitaxial films, which typically relax below $2\%$ strain, but may be accessible in free-standing membranes. 

Strain gradient in membranes can also tune topological states via pseudo magnetic fields. While homogeneous strains have zero gauge field, spatially varying strains in materials produce a pseudo magnetic field \textbf{B}=$\nabla\times$\textbf{A} \cite{jiang2017visualizing,kang2021pseudo}. A previous study showed that the pseudo field created by dislocation arrays can flatten the bands near the Dirac points to create helical surface states \cite{tang2014strain}. Membranes provide an alternative path to more controllably create inhomogeneous strain fields and their associated pseudo magnetic fields, borrowing techniques that have been developed for inducing pseudo B fields in graphene \cite{jiang2017visualizing,kang2021pseudo}. The pseudo magnetic fields are also powerful for tuning the k-space spacing between Weyl nodes \cite{armitage2018weyl}, which act as sources and sinks of Berry curvature.

\section{Why now?}

The new science and strain engineering of single crystalline membranes is driven by recent advances in membrane synthesis, demonstrations of extreme and tunable strains, and integrated computational modelling from atomistic to mesoscale.

\subsection{New membrane synthesis}

\begin{figure}[ht]
    \centering
    \includegraphics[width=0.5\textwidth]{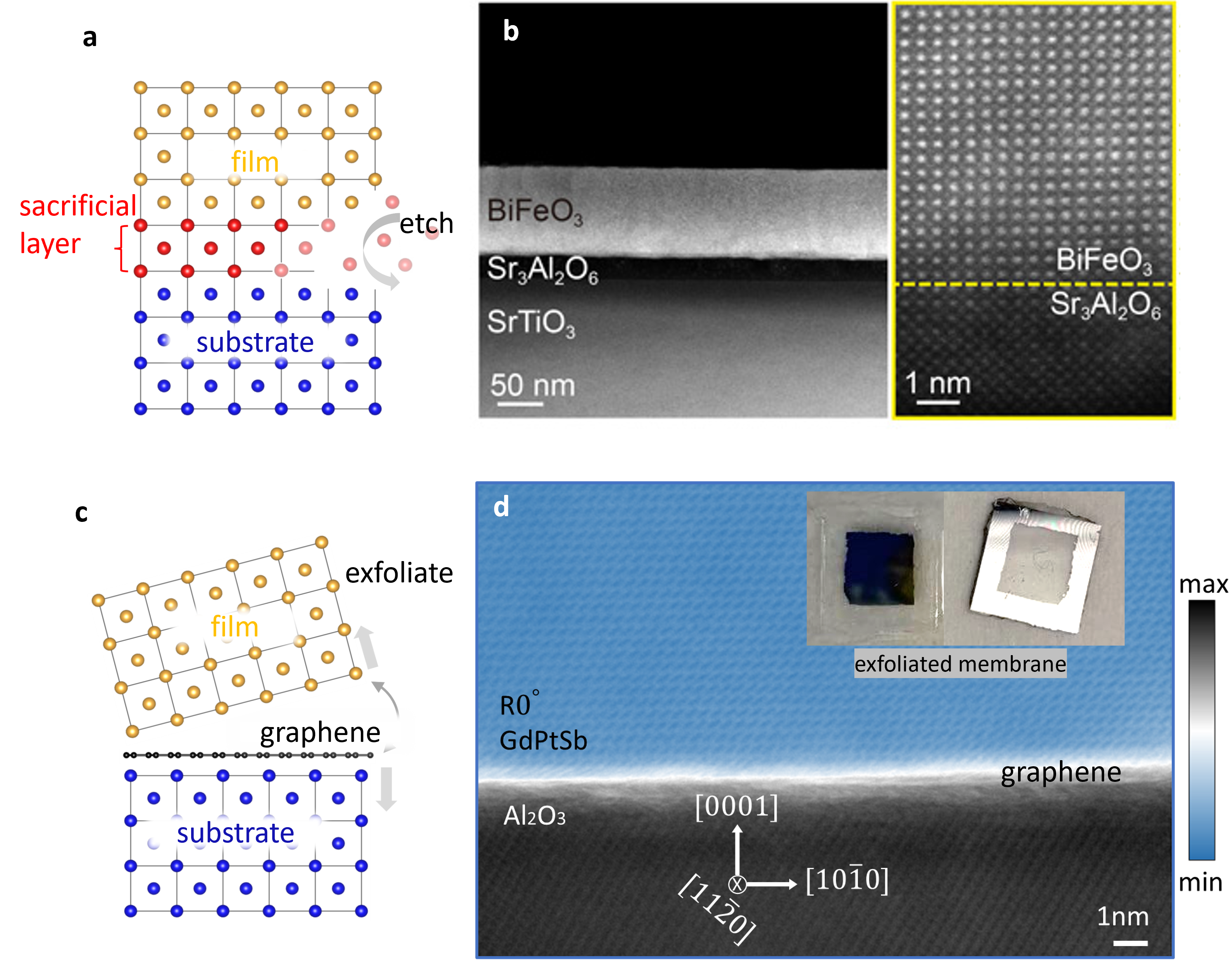}
    \caption{Synthesis of single-crystalline membranes. (a) Epitaxial etch release. (b) Transmission electron microscopy (TEM) image of a BiFeO$_3$ (BFO) film grown on a Sr$_3$Al$_2$O$_6$ (SAO) sacrificial etch layer. From Peng et. al.,  Sci. Adv., 6, aba5847 (2020) (Ref. \cite{peng2020phase}). Reproduced with permission from AAAS under Creative Commons License. (c) Remote epitaxy and exfoliation from graphene. (d) Epitaxy of GdPtSb on graphene/Al$_2$O$_3$(0001), reproduced under Creative Commons License from Ref. \cite{du2021epitaxy}. Inset photos show the GdPtSb membrane and the graphene/Al$_2$O$_3$ substrate after exfoliation.}
    \label{synthesis}
\end{figure}

Epitaxial growth and release from a sacrificial etch layer is a leading membrane synthesis strategy (Fig. \ref{synthesis}a,b). This approach was first developed for semiconductor membranes, including SiGe membranes by etching the oxide from silicon on insulator (SOI) \cite{roberts2006elastically}, and GaAs/AlAs membranes by selective etches for GaAs or AlAs layers \cite{cheng2013epitaxial}. It has been extended to other materials that lattice match to semiconductors, including the shape memory alloy Ni$_2$MnGa fabricated via epitaxial growth on AlGaAs and subsequent etching \cite{dong2004shape}. 

New water soluble oxide layers enable the release of free-standing perovskite transition metal oxide membranes. These release layers include (Ca,Sr,Ba)$_3$Al$_2$O$_6$, which allows the lattice parameter to be tuned from 3.819 \AA\ to 4.124 \AA\ \cite{ji2019freestanding, hong2020extreme, chiabrera2022freestanding}, SrVO$_3$ \cite{bourlier2020transfer}, and BaO \cite{takahashi2020sacrificial}. These layers are typically grown by pulsed laser deposition (PLD)\cite{hong2020extreme} or molecular beam epitaxy (MBE) \cite{ji2019freestanding}. A significant challenge for epitaxial etch release is that not all materials combinations have selective etch chemistries that can etch the lattice matched release layer without damaging the membrane layer. 

Remote epitaxy and exfoliation provide an etch-free alternative (Fig. \ref{synthesis}c,d). In this approach, an epitaxial film is grown on a graphene (or other 2D material) covered substrate \cite{kim2017remote}. Epitaxial registry between film and substrate is thought to occur via remote interactions that permeate through graphene \cite{kim2017remote, kong2018polarity}, although a pinhole-seeded mechanism can also produce exfoliatable membranes \cite{manzo2022pinhole}. The weak van der Waals interactions of graphene allow film exfoliation to produce a freestanding membrane, similar to exfoliation of vdW materials like graphene and transition metal dichalcogenides. First demonstrated for the compound semiconductors \cite{kim2017remote}, growth and exfoliation from graphene has been demonstrated for transition metal oxides \cite{kum2020heterogeneous, yoon2022free}, halide perovskites \cite{jiang2019carrier}, simple metals \cite{lu2018remote}, and Heusler compounds \cite{du2021epitaxy, du2022controlling}.

Several challenges exist for remote epitaxy. First, the quality of remote epitaxial film growth and ability to exfoliate depend on the quality of the starting 2D material covered substrate. In most cases, this starting surface is prepared by layer transfer because graphene and other 2D materials cannot be grown directly on arbitrary substrates. This transfer can introduce wrinkles, tears, and interfacial contaminants that introduce defects in the subsequent membrane growth \cite{manzo2022pinhole}, and in extreme cases can affect the ability to exfoliate \cite{kim2021role, kim2021impact}. A cleaner alternative strategy is to use graphene directly grown on the substrate of interest. Recently, epitaxial BaTiO$_3$ membranes were grown on graphene/Ge (110) \cite{dai2022highly}, where the graphene was grown directly on Ge. Further advancements in remote epitaxy may require the development of graphene growth directly on new substrates of interest.

A second challenge is that the atomic-scale mechanisms for remote epitaxy remain unclear. Clear experimental evidence for a remote mechanism remains elusive. In most experiments, the primary evidence for a remote mechanism is that the films are epitaxial to the underlying substrate (rather than to graphene) and can be exfoliated. Recent in-situ surface science measurements, however, demonstrate that a pinhole-seeded lateral epitaxy mechanism can also produce epitaxial, exfoliatable membranes \cite{manzo2022pinhole}. In this growth mode, few nanometer diameter pinholes in the graphene serve as sites for selective nucleation at the substrate, followed by lateral overgrowth and coalescence of a continuous film. Since the pinholes are small and sparse, membranes can still be exfoliated. Moreover, the pinholes are easy to overlook because they do not appear after the graphene transfer step.  Instead they only appear immediately prior to film growth because they are created by interfacial oxide desorption at pre-growth sample annealing temperatures.

Careful microscopic measurements at multiple steps during the growth process are required to understand the growth mechanisms on graphene. The development of graphene grown directly on substrates of interest, e.g. graphene on Ge, avoids the interfacial oxide-induced pinholes and may allow the intrinsic mechanisms for remote epitaxy to be tested. Alternative forms of evidence may also shed light on the mechanisms: for GdPtSb films grown on graphene/Al$_2$O$_3$ (0001), a $30^\circ$  rotated superstructure forms that cannot be explained by pinholes \cite{du2022controlling}. Is this superstructure evidence for an intrinsic remote epitaxy mechanism? A microscopic understanding of the mechanisms, whether intrinsic remote epitaxy or extrinsic pinholes, is required to understand the limits and new applications for epitaxy and exfoliation from graphene.

\subsection{Extreme strain manipulation}

\begin{figure*}[ht]
    \centering
    \includegraphics[width=0.8\textwidth]{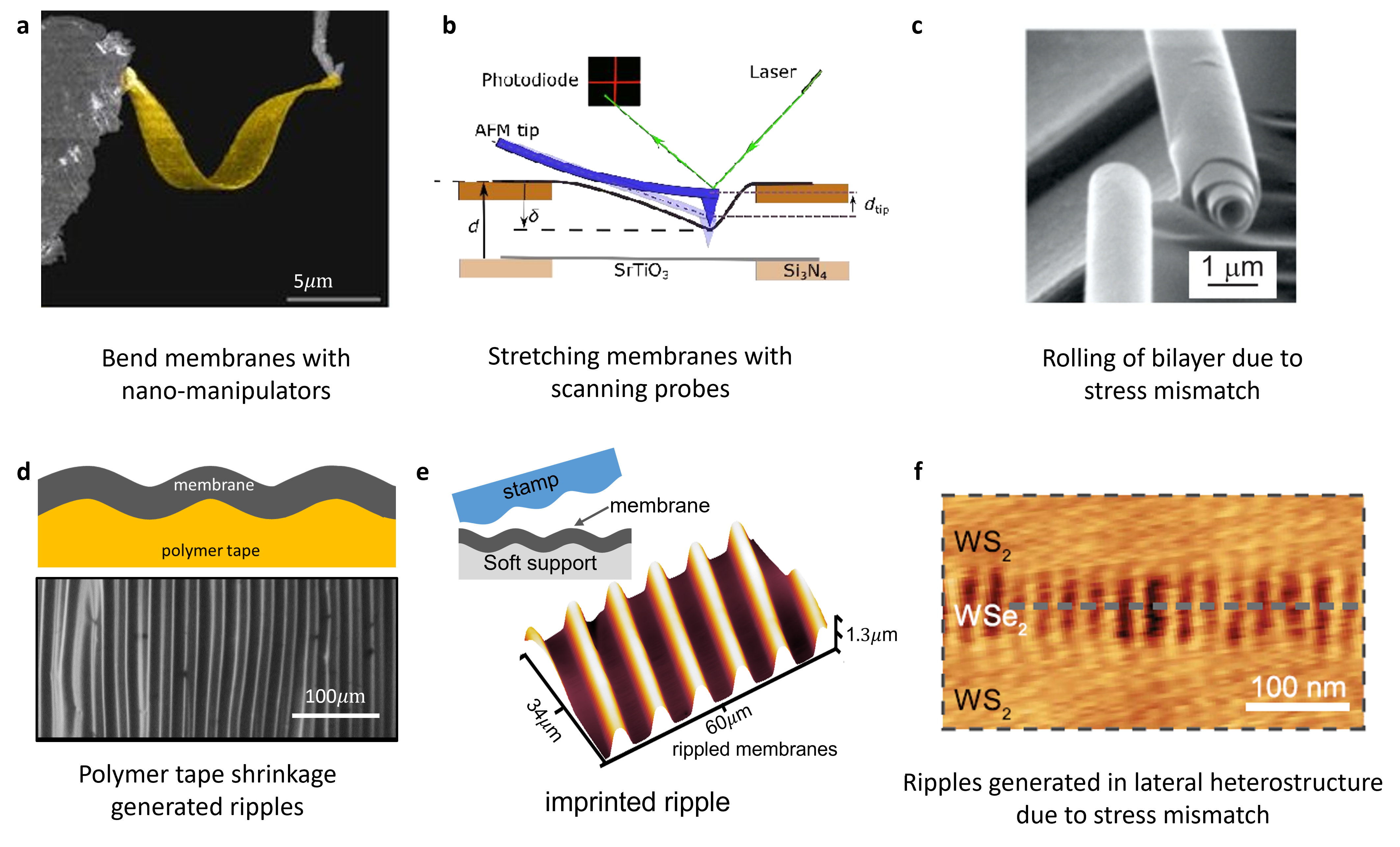}
    \caption{Methods of generating strain gradients. (a) Bending membrane ribbons with nano-manipulator. From Peng, et. al. Science Advances 6, eaba5847 (2020) (Ref \cite{peng2020phase}). Reprinted under Creative Commons License. (b) Stretching membranes by AFM tips. Reprinted (adapted) with permission from V. Harbola et. al., Nano Lett. 21, 6, 2470–2475 (2021) (Ref. \cite{harbola2021strain}). Copyright 2021 American Chemical Society.
   (c) Rolling up of SrTiO$_3$/Si/SiGe membrane via strain relaxation. From Prakash, et. al., Small 18, 2105424 (2022) (Ref. \cite{prakash2022reconfiguration}). Reprinted with permission from WILEY. (d) Transferring membranes to pre-strained polymer tape and generating ripples by expanding and shrinking of the tape. (e) Rippling membranes by transferring them to soft support and imprinting patterns to membranes with stamps. (f) Generating ripples in in-plane heterostructures via strain relaxation. From Xie, et. al., Science 359, 1131 (2018) (Ref. \cite{xie2018coherent}). Reprinted with permission from AAAS.}
    \label{bending and rippling}
\end{figure*}

Released membranes enable the application of extreme strains. To date, strain is typically applied via top-down methods. Using micropositioners, strains of 8\% have been demonstrated in few nanometer thick (La,Ca)MnO$_3$ membranes in tension \cite{hong2020extreme}, and $5.4\%$ for BiFeO$_3$ \cite{peng2020phase} and $\sim 10\%$ for BaTiO$_3$ membranes \cite{dong2019super} in bending. A flexible polymer handle can aide in the handling of ultrathin membranes, and the use of polymers handles cooled below the glass transition temperature can lock in the desired strain state \cite{hong2020extreme}. 

Strain gradients can be produced by bending and rippling. Methods include local bending using a scanning probe or micropositioners \cite{lee2020flexoelectricity, harbola2021strain}, Fig. \ref{bending and rippling} (a,b), rippling via lateral compression on a polymer handle \cite{du2021epitaxy, cai2022enhanced}, and transferring membranes to a patterned surface \cite{kang2021pseudo}, Fig. \ref{bending and rippling} (d,e). Local bending by micropositioners have demonstrated elastic recoverable 180 degree bends with 1 micon radius of curvature, as is shown in Fig. \ref{bending and rippling} (a) \cite{peng2020phase}.

Bottom-up strategies provide new opportunities for fine strain gradient control. Strain sharing bilayers, in which one layer is compressive and the other is tensile, spontaneously roll up into nanotubes upon release. This strategy has been successfully implemented to make semiconductor nanotubes \cite{schmidt2001thin, huang2005nanomechanical} and curved oxide membranes \cite{prakash2022reconfiguration} (Fig. \ref{bending and rippling} (c)). Another strategy is spontaneous rippling in lattice-mismatched lateral heterostuctures. First implemented for WS$_2$/WSe$_2$ heterostructures, these materials relax via rippling out of plane due to the weak van der Waals interaction with the substrate \cite{xie2018coherent}, as is shown in Fig. \ref{bending and rippling} (f). We envision similar lateral heterostructures of non vdW membranes, grown by remote epitaxy on graphene, may experience out of plane rippling.

\subsection{Why can membranes sustain much larger strains than clamped films or bulk materials?} 

We offer several possible reasons, based on surface science \cite{zangwill1988physics} and the mechanics of 1D metallic whiskers \cite{sears1954elastic, herring1952elastic, brenner1956tensile} and semiconductor and metallic nanowires \cite{wang2011super, wang2017mechanical, yue2011approaching, li2021superelastic,deng2019hierarchically}.

First, membranes are not clamped to a rigid substrate. In epitaxial films, dislocations form when the strain energy exceeds the energy cost to form a misfit dislocation at the film/substrate interface. This criterion, which can be expressed in terms of an energy balance (People and Bean \cite{people1985calculation}, van der Merwe \cite{van1963crystal}) or a force balance (Matthews and Blakeslee \cite{matthews1974defects}), typically limits strains to $\sim 2$\%. Otherwise a film relaxes at a critical thickness below one unit cell. For a free-standing membrane, there is no interfacial bonding between film and substrate to create a misfit dislocation. Thus dislocations must nucleate from the bulk or from the top or bottom surface.

Second, ultrathin membranes are dominated by their surfaces. At surface, atoms have decreased local coordination and increased degrees of freedom for relaxation compared to bulk. In response to external stresses, surface atoms can relax out-of-plane or reconstruct in-plane. Surface contributions \cite{muller2004elastic, liu2009size} are invoked to explain the elasticity of few nanometer diameter nanowires, which can also sustain elastic strains of order $\sim 10 \%$ \cite{wang2011super, wang2017mechanical, yue2011approaching}. Similar arguments may explain why a 6 nm thick (La,Ca)MnO$_3$ membranes can sustain 8\% elastic strain, whereas thicker membranes ($> 20$ nm) undergo fracture below 2\% strain \cite{hong2020extreme}. 

Interestingly, novel phase transitions and domain reorientations have been observed by transmission electron microscopy in bent ultrathin membranes of the ferroelectric materials BaTiO$_3$ \cite{dong2019super} and BiFeO$_3$ \cite{peng2020phase}, and a continuous face centered cubic to body centered tetragonal transition has been detected in few nanometer diameter Cu nanowires \cite{wang2013situ}. These studies indicate that elastic deformations within the \textit{interior} of a thin membrane, and not just within the few surface layers, can be different than the bulk. Further microscopic studies are needed in order to understand the relaxations and reconstructions at the surface and near surface region of strained ultrathin membranes.

Third, the mechanisms for generation, motion, and pinning of defects are length scale dependent \cite{uchic2004sample}. Activation and suppression of these mechanisms has been invoked to explain the size-dependent elastic properties of few micron diameter metallic whiskers \cite{sears1954elastic,herring1952elastic, brenner1956tensile} and micropillars \cite{uchic2004sample}. Similar arguments may describe the mechanics of membranes at intermediate thicknesses of tens to hundereds of nanometers.

\subsection{New developments and challenges in modeling.} 

An accurate modeling and prediction of the physical properties of strained membranes requires theory and computation at multiple scales. Of central importance is the accurate treatment of the spatially inhomogeneous strain (e.g., strain gradient), which has been challenging to address through first-principles density functional theories (DFT) calculations. This is because inhomogeneous strain often creates non-periodic crystal structure (incommensurate lattice distortion) yet the supercell used in DFT calculation often needs to be periodic.  Thanks to the recent advances in the density-functional perturbation theory (DFPT), it is now possible to accurately compute the microscopic response (both linear and non-linear) of a system to an arbitrary inhomogeneous strain. Perhaps the most prominent example is the development of first-principles theory of flexoelectricity \cite{stengel2017first} and its application to compute the flexoelectric tensor \cite{dreyer2018current, schiaffino2019metric, royo2019first, springolo2021direct, royo2022lattice}, which can then be utilized to inform the mesoscale/continuum materials modeling \cite{springolo2021direct}.   

Despite these exciting developments, significant challenges still remain. For example, the properties of strained membrane, like most practical materials, depend on the formation and evolution of mesoscale patterns (e.g., magnetic/ferroelectric/ferroelastic domains, electronic phase separation) at finite temperature, which go beyond the capability of conventional DFT calculations. However, research into the prediction of mesoscale pattern formation under extreme strain condition is still at its early stage, with many open questions remain. Take the  ferroelectrics as an example, large bending can significantly change the bandgap of the domain wall \cite{eliseev2012conductivity} and hence lead to redistribution of the ionic and electronic defects and even an insulator-to-metal transition \cite{stolichnov2015bent}. How does the strain-induced ionic/electronic defect redistribution interact with the domain structure evolution under extreme strain condition \cite{deng2019hierarchically}? How does the defect distribution influence the strain-induced ferroelectric/ferroelastic phase transition? How to disentangle the contribution of flexoelectricity, piezoelectricity, and electrostriction to the mesoscale pattern formation?  In addition to these fundamental challenges, there also exist technical challenges in different computational methods. 

Modern atomistic methods such as effective Hamiltonian-based methods\cite{zhong1994phase, bellaiche2000finite, lai2006electric,nahas2020inverse} and second-principles calculations \cite{wojdel2013first, wojdel2014ferroelectric,garcia2016second} can predict the mesoscale pattern formation with atom-resolved spatial resolution, and permits taking input directly from DFT calculations without the need of parameterization. However, it is still challenging to consider the realistic mechanical boundary conditions for the application of strain and strain gradient (Fig. \ref{bending and rippling}) and their application to practical-sized (e.g., hundreds of micrometers to millimeters) materials systems currently would consume too much computational resources to be realistic. 

Mesoscale materials modeling methods such as phase-field modeling cannot predict pattern formation and evolution at the scale below one unit cell, but can conveniently consider the complexity arising from the actual mechanical boundary condition upon the application of strain (gradient) \cite{deng2019hierarchically, li2021superelastic, park2022light, peng2020phase, dong2019super}, and incorporate the role of 0D (point defects such as oxygen vacancies \cite{deng2019hierarchically}), 1D (dislocations \cite{zhuo2022anisotropic}), 2D (grain boundaries \cite{dai2021graph, choudhury2007effect, hu2019phase}), and 3D (e.g., precipitates \cite{zhao2021precipitation} and cracks \cite{liu2018electrically}). In particular, the phase-field model has the additional versatility of modeling the formation and co-evolution of different types of coupled patterns, for example, the coupled magnetic and structural domains \cite{ohmer2022phase, huang2015phase, wu2011phase}.  With input from ab initio and/or experimental measurements, the predicted mesoscale patterns can often be utilized for a side-by-side comparison to experiments for not only understanding and interpreting the results, but also provide insights into how to access these patterns and manipulate them for realizing exotic phenomena or enhanced responses \cite{qiu2020transparent, pan2019ultrahigh, pan2021ultrahigh}.

\section{Outlook: beyond static strains}

Large strains and strain gradients provide unique opportunities for inducing new properties in membranes of quantum materials. This Perspective highlighted static strain tuning of magnetism, superconductivity, ferroelectricity, and topological states.

Exciting opportunities also lie in dynamic and nonequilibirum properties. Nonlinear phononics, in which ultrafast optical pulses resonantly excite phonon modes, is a powerful approach for revealing nonequilibrium properties that arise from photon-phonon-spin or photon-phonon-electron couplings. Examples include ultrafast antiferromagnetic-ferrimagnetic switching \cite{disa2020polarizing}, metastable ferroelectricity \cite{nova2019metastable}, and possible nonequilibrium superconductivity \cite{budden2021evidence}. The general applicability of nonlinear phononics, however, is limited since these complex couplings are often weak, difficult to tune, and difficult to apply beyond a narrow set of materials that obey the required symmetry constraints. We anticipate the strong symmetry-breaking strains and strain gradients in membranes may solve this challenge, by enhancing the quasiparticle coupling strengths via strain, and breaking symmetries to activate new phonon modes for resonant excitation. The absence of substrate clamping is also beneficial since larger amplitude lattice vibrations can be accessed. Strain and strain gradients, both in static and dynamic forms, provide power tuning knobs for unleashing hidden properties in quantum materials membranes.

\section{Acknowledgements}

We thank Cyrus Dreyer, Jun Xiao, Daniel Rhodes, Ying Wang, and Uwe Bergmann for discussions. JKK and DD acknowledge the Air Force Office of Scientific Research (FA9550-21-0127) and the National Science Foundation (DMR-1752797). All authors acknowledge the National Science Foundation through the University of Wisconsin Materials Research Science and Engineering Center (MRSEC) Grant No. DMR-1720415.

\bibliographystyle{apsrev}
\bibliography{ref}

\begin{thebibliography}{134}
\expandafter\ifx\csname natexlab\endcsname\relax\def\natexlab#1{#1}\fi
\expandafter\ifx\csname bibnamefont\endcsname\relax
  \def\bibnamefont#1{#1}\fi
\expandafter\ifx\csname bibfnamefont\endcsname\relax
  \def\bibfnamefont#1{#1}\fi
\expandafter\ifx\csname citenamefont\endcsname\relax
  \def\citenamefont#1{#1}\fi
\expandafter\ifx\csname url\endcsname\relax
  \def\url#1{\texttt{#1}}\fi
\expandafter\ifx\csname urlprefix\endcsname\relax\def\urlprefix{URL }\fi
\providecommand{\bibinfo}[2]{#2}
\providecommand{\eprint}[2][]{\url{#2}}

\bibitem[{\citenamefont{Ruf et~al.}(2021)\citenamefont{Ruf, Paik, Schreiber,
  Nair, Miao, Kawasaki, Nelson, Faeth, Lee, Goodge et~al.}}]{ruf2021strain}
\bibinfo{author}{\bibfnamefont{J.~P.} \bibnamefont{Ruf}},
  \bibinfo{author}{\bibfnamefont{H.}~\bibnamefont{Paik}},
  \bibinfo{author}{\bibfnamefont{N.~J.} \bibnamefont{Schreiber}},
  \bibinfo{author}{\bibfnamefont{H.~P.} \bibnamefont{Nair}},
  \bibinfo{author}{\bibfnamefont{L.}~\bibnamefont{Miao}},
  \bibinfo{author}{\bibfnamefont{J.~K.} \bibnamefont{Kawasaki}},
  \bibinfo{author}{\bibfnamefont{J.~N.} \bibnamefont{Nelson}},
  \bibinfo{author}{\bibfnamefont{B.~D.} \bibnamefont{Faeth}},
  \bibinfo{author}{\bibfnamefont{Y.}~\bibnamefont{Lee}},
  \bibinfo{author}{\bibfnamefont{B.~H.} \bibnamefont{Goodge}},
  \bibnamefont{et~al.}, \bibinfo{journal}{Nature Communications}
  \textbf{\bibinfo{volume}{12}}, \bibinfo{pages}{1} (\bibinfo{year}{2021}).

\bibitem[{\citenamefont{Schlom et~al.}(2007)\citenamefont{Schlom, Chen, Eom,
  Rabe, Streiffer, and Triscone}}]{schlom2007strain}
\bibinfo{author}{\bibfnamefont{D.~G.} \bibnamefont{Schlom}},
  \bibinfo{author}{\bibfnamefont{L.-Q.} \bibnamefont{Chen}},
  \bibinfo{author}{\bibfnamefont{C.-B.} \bibnamefont{Eom}},
  \bibinfo{author}{\bibfnamefont{K.~M.} \bibnamefont{Rabe}},
  \bibinfo{author}{\bibfnamefont{S.~K.} \bibnamefont{Streiffer}},
  \bibnamefont{and} \bibinfo{author}{\bibfnamefont{J.-M.}
  \bibnamefont{Triscone}}, \bibinfo{journal}{Annual Review of Materials
  Research} \textbf{\bibinfo{volume}{37}}, \bibinfo{pages}{589}
  (\bibinfo{year}{2007}).

\bibitem[{\citenamefont{Song et~al.}(2020)\citenamefont{Song, Xiong, Zhang,
  Nie, and Li}}]{song2020strain}
\bibinfo{author}{\bibfnamefont{X.}~\bibnamefont{Song}},
  \bibinfo{author}{\bibfnamefont{C.}~\bibnamefont{Xiong}},
  \bibinfo{author}{\bibfnamefont{F.}~\bibnamefont{Zhang}},
  \bibinfo{author}{\bibfnamefont{Y.}~\bibnamefont{Nie}}, \bibnamefont{and}
  \bibinfo{author}{\bibfnamefont{Y.}~\bibnamefont{Li}},
  \bibinfo{journal}{Materials Letters} \textbf{\bibinfo{volume}{259}},
  \bibinfo{pages}{126914} (\bibinfo{year}{2020}).

\bibitem[{\citenamefont{Matthews and Blakeslee}(1974)}]{matthews1974defects}
\bibinfo{author}{\bibfnamefont{J.}~\bibnamefont{Matthews}} \bibnamefont{and}
  \bibinfo{author}{\bibfnamefont{A.}~\bibnamefont{Blakeslee}},
  \bibinfo{journal}{Journal of Crystal growth} \textbf{\bibinfo{volume}{27}},
  \bibinfo{pages}{118} (\bibinfo{year}{1974}).

\bibitem[{\citenamefont{People and Bean}(1985)}]{people1985calculation}
\bibinfo{author}{\bibfnamefont{R.}~\bibnamefont{People}} \bibnamefont{and}
  \bibinfo{author}{\bibfnamefont{J.}~\bibnamefont{Bean}},
  \bibinfo{journal}{Applied Physics Letters} \textbf{\bibinfo{volume}{47}},
  \bibinfo{pages}{322} (\bibinfo{year}{1985}).

\bibitem[{\citenamefont{Van Der~Merwe}(1963)}]{van1963crystal}
\bibinfo{author}{\bibfnamefont{J.~H.} \bibnamefont{Van Der~Merwe}},
  \bibinfo{journal}{Journal of Applied Physics} \textbf{\bibinfo{volume}{34}},
  \bibinfo{pages}{123} (\bibinfo{year}{1963}).

\bibitem[{\citenamefont{Hong et~al.}(2020)\citenamefont{Hong, Gu, Verma,
  Harbola, Wang, Lu, Vailionis, Hikita, Pentcheva, Rondinelli
  et~al.}}]{hong2020extreme}
\bibinfo{author}{\bibfnamefont{S.~S.} \bibnamefont{Hong}},
  \bibinfo{author}{\bibfnamefont{M.}~\bibnamefont{Gu}},
  \bibinfo{author}{\bibfnamefont{M.}~\bibnamefont{Verma}},
  \bibinfo{author}{\bibfnamefont{V.}~\bibnamefont{Harbola}},
  \bibinfo{author}{\bibfnamefont{B.~Y.} \bibnamefont{Wang}},
  \bibinfo{author}{\bibfnamefont{D.}~\bibnamefont{Lu}},
  \bibinfo{author}{\bibfnamefont{A.}~\bibnamefont{Vailionis}},
  \bibinfo{author}{\bibfnamefont{Y.}~\bibnamefont{Hikita}},
  \bibinfo{author}{\bibfnamefont{R.}~\bibnamefont{Pentcheva}},
  \bibinfo{author}{\bibfnamefont{J.~M.} \bibnamefont{Rondinelli}},
  \bibnamefont{et~al.}, \bibinfo{journal}{Science}
  \textbf{\bibinfo{volume}{368}}, \bibinfo{pages}{71} (\bibinfo{year}{2020}).

\bibitem[{\citenamefont{Dong et~al.}(2019)\citenamefont{Dong, Li, Yao, Zhou,
  Zhang, Han, Luo, Yao, Peng, Hu et~al.}}]{dong2019super}
\bibinfo{author}{\bibfnamefont{G.}~\bibnamefont{Dong}},
  \bibinfo{author}{\bibfnamefont{S.}~\bibnamefont{Li}},
  \bibinfo{author}{\bibfnamefont{M.}~\bibnamefont{Yao}},
  \bibinfo{author}{\bibfnamefont{Z.}~\bibnamefont{Zhou}},
  \bibinfo{author}{\bibfnamefont{Y.-Q.} \bibnamefont{Zhang}},
  \bibinfo{author}{\bibfnamefont{X.}~\bibnamefont{Han}},
  \bibinfo{author}{\bibfnamefont{Z.}~\bibnamefont{Luo}},
  \bibinfo{author}{\bibfnamefont{J.}~\bibnamefont{Yao}},
  \bibinfo{author}{\bibfnamefont{B.}~\bibnamefont{Peng}},
  \bibinfo{author}{\bibfnamefont{Z.}~\bibnamefont{Hu}}, \bibnamefont{et~al.},
  \bibinfo{journal}{Science} \textbf{\bibinfo{volume}{366}},
  \bibinfo{pages}{475} (\bibinfo{year}{2019}).

\bibitem[{\citenamefont{Du et~al.}(2021)\citenamefont{Du, Manzo, Zhang,
  Saraswat, Genser, Rabe, Voyles, Arnold, and Kawasaki}}]{du2021epitaxy}
\bibinfo{author}{\bibfnamefont{D.}~\bibnamefont{Du}},
  \bibinfo{author}{\bibfnamefont{S.}~\bibnamefont{Manzo}},
  \bibinfo{author}{\bibfnamefont{C.}~\bibnamefont{Zhang}},
  \bibinfo{author}{\bibfnamefont{V.}~\bibnamefont{Saraswat}},
  \bibinfo{author}{\bibfnamefont{K.~T.} \bibnamefont{Genser}},
  \bibinfo{author}{\bibfnamefont{K.~M.} \bibnamefont{Rabe}},
  \bibinfo{author}{\bibfnamefont{P.~M.} \bibnamefont{Voyles}},
  \bibinfo{author}{\bibfnamefont{M.~S.} \bibnamefont{Arnold}},
  \bibnamefont{and} \bibinfo{author}{\bibfnamefont{J.~K.}
  \bibnamefont{Kawasaki}}, \bibinfo{journal}{Nature communications}
  \textbf{\bibinfo{volume}{12}}, \bibinfo{pages}{1} (\bibinfo{year}{2021}).

\bibitem[{\citenamefont{Kim et~al.}(2017)\citenamefont{Kim, Cruz, Lee, Alawode,
  Choi, Song, Johnson, Heidelberger, Kong, Choi et~al.}}]{kim2017remote}
\bibinfo{author}{\bibfnamefont{Y.}~\bibnamefont{Kim}},
  \bibinfo{author}{\bibfnamefont{S.~S.} \bibnamefont{Cruz}},
  \bibinfo{author}{\bibfnamefont{K.}~\bibnamefont{Lee}},
  \bibinfo{author}{\bibfnamefont{B.~O.} \bibnamefont{Alawode}},
  \bibinfo{author}{\bibfnamefont{C.}~\bibnamefont{Choi}},
  \bibinfo{author}{\bibfnamefont{Y.}~\bibnamefont{Song}},
  \bibinfo{author}{\bibfnamefont{J.~M.} \bibnamefont{Johnson}},
  \bibinfo{author}{\bibfnamefont{C.}~\bibnamefont{Heidelberger}},
  \bibinfo{author}{\bibfnamefont{W.}~\bibnamefont{Kong}},
  \bibinfo{author}{\bibfnamefont{S.}~\bibnamefont{Choi}}, \bibnamefont{et~al.},
  \bibinfo{journal}{Nature} \textbf{\bibinfo{volume}{544}},
  \bibinfo{pages}{340} (\bibinfo{year}{2017}).

\bibitem[{\citenamefont{Kum et~al.}(2020)\citenamefont{Kum, Lee, Kim,
  Lindemann, Kong, Qiao, Chen, Irwin, Lee, Xie et~al.}}]{kum2020heterogeneous}
\bibinfo{author}{\bibfnamefont{H.~S.} \bibnamefont{Kum}},
  \bibinfo{author}{\bibfnamefont{H.}~\bibnamefont{Lee}},
  \bibinfo{author}{\bibfnamefont{S.}~\bibnamefont{Kim}},
  \bibinfo{author}{\bibfnamefont{S.}~\bibnamefont{Lindemann}},
  \bibinfo{author}{\bibfnamefont{W.}~\bibnamefont{Kong}},
  \bibinfo{author}{\bibfnamefont{K.}~\bibnamefont{Qiao}},
  \bibinfo{author}{\bibfnamefont{P.}~\bibnamefont{Chen}},
  \bibinfo{author}{\bibfnamefont{J.}~\bibnamefont{Irwin}},
  \bibinfo{author}{\bibfnamefont{J.~H.} \bibnamefont{Lee}},
  \bibinfo{author}{\bibfnamefont{S.}~\bibnamefont{Xie}}, \bibnamefont{et~al.},
  \bibinfo{journal}{Nature} \textbf{\bibinfo{volume}{578}}, \bibinfo{pages}{75}
  (\bibinfo{year}{2020}).

\bibitem[{\citenamefont{Du et~al.}(2022)\citenamefont{Du, Jung, Manzo, LaDuca,
  Zheng, Su, Saraswat, McChesney, Arnold, and Kawasaki}}]{du2022controlling}
\bibinfo{author}{\bibfnamefont{D.}~\bibnamefont{Du}},
  \bibinfo{author}{\bibfnamefont{T.}~\bibnamefont{Jung}},
  \bibinfo{author}{\bibfnamefont{S.}~\bibnamefont{Manzo}},
  \bibinfo{author}{\bibfnamefont{Z.}~\bibnamefont{LaDuca}},
  \bibinfo{author}{\bibfnamefont{X.}~\bibnamefont{Zheng}},
  \bibinfo{author}{\bibfnamefont{K.}~\bibnamefont{Su}},
  \bibinfo{author}{\bibfnamefont{V.}~\bibnamefont{Saraswat}},
  \bibinfo{author}{\bibfnamefont{J.}~\bibnamefont{McChesney}},
  \bibinfo{author}{\bibfnamefont{M.~S.} \bibnamefont{Arnold}},
  \bibnamefont{and} \bibinfo{author}{\bibfnamefont{J.~K.}
  \bibnamefont{Kawasaki}}, \bibinfo{journal}{Nano Letters}
  \textbf{\bibinfo{volume}{22}}, \bibinfo{pages}{8647} (\bibinfo{year}{2022}).

\bibitem[{\citenamefont{Yoon et~al.}(2022)\citenamefont{Yoon, Truttmann, Liu,
  Matthews, Choo, Su, Saraswat, Manzo, Arnold, Bowden et~al.}}]{yoon2022free}
\bibinfo{author}{\bibfnamefont{H.}~\bibnamefont{Yoon}},
  \bibinfo{author}{\bibfnamefont{T.~K.} \bibnamefont{Truttmann}},
  \bibinfo{author}{\bibfnamefont{F.}~\bibnamefont{Liu}},
  \bibinfo{author}{\bibfnamefont{B.~E.} \bibnamefont{Matthews}},
  \bibinfo{author}{\bibfnamefont{S.}~\bibnamefont{Choo}},
  \bibinfo{author}{\bibfnamefont{Q.}~\bibnamefont{Su}},
  \bibinfo{author}{\bibfnamefont{V.}~\bibnamefont{Saraswat}},
  \bibinfo{author}{\bibfnamefont{S.}~\bibnamefont{Manzo}},
  \bibinfo{author}{\bibfnamefont{M.~S.} \bibnamefont{Arnold}},
  \bibinfo{author}{\bibfnamefont{M.~E.} \bibnamefont{Bowden}},
  \bibnamefont{et~al.}, \bibinfo{journal}{arXiv preprint arXiv:2206.09094}
  (\bibinfo{year}{2022}).

\bibitem[{\citenamefont{Dai et~al.}(2022)\citenamefont{Dai, Zhao, Li, Chen,
  Zhai, Xue, Di, Feng, Sun, Luo et~al.}}]{dai2022highly}
\bibinfo{author}{\bibfnamefont{L.}~\bibnamefont{Dai}},
  \bibinfo{author}{\bibfnamefont{J.}~\bibnamefont{Zhao}},
  \bibinfo{author}{\bibfnamefont{J.}~\bibnamefont{Li}},
  \bibinfo{author}{\bibfnamefont{B.}~\bibnamefont{Chen}},
  \bibinfo{author}{\bibfnamefont{S.}~\bibnamefont{Zhai}},
  \bibinfo{author}{\bibfnamefont{Z.}~\bibnamefont{Xue}},
  \bibinfo{author}{\bibfnamefont{Z.}~\bibnamefont{Di}},
  \bibinfo{author}{\bibfnamefont{B.}~\bibnamefont{Feng}},
  \bibinfo{author}{\bibfnamefont{Y.}~\bibnamefont{Sun}},
  \bibinfo{author}{\bibfnamefont{Y.}~\bibnamefont{Luo}}, \bibnamefont{et~al.},
  \bibinfo{journal}{Nature Communications} \textbf{\bibinfo{volume}{13}},
  \bibinfo{pages}{1} (\bibinfo{year}{2022}).

\bibitem[{\citenamefont{Si et~al.}(2016)\citenamefont{Si, Sun, and
  Liu}}]{si2016strain}
\bibinfo{author}{\bibfnamefont{C.}~\bibnamefont{Si}},
  \bibinfo{author}{\bibfnamefont{Z.}~\bibnamefont{Sun}}, \bibnamefont{and}
  \bibinfo{author}{\bibfnamefont{F.}~\bibnamefont{Liu}},
  \bibinfo{journal}{Nanoscale} \textbf{\bibinfo{volume}{8}},
  \bibinfo{pages}{3207} (\bibinfo{year}{2016}).

\bibitem[{\citenamefont{Peng et~al.}(2020{\natexlab{a}})\citenamefont{Peng,
  Chen, Fan, Srolovitz, and Lei}}]{peng2020strain}
\bibinfo{author}{\bibfnamefont{Z.}~\bibnamefont{Peng}},
  \bibinfo{author}{\bibfnamefont{X.}~\bibnamefont{Chen}},
  \bibinfo{author}{\bibfnamefont{Y.}~\bibnamefont{Fan}},
  \bibinfo{author}{\bibfnamefont{D.~J.} \bibnamefont{Srolovitz}},
  \bibnamefont{and} \bibinfo{author}{\bibfnamefont{D.}~\bibnamefont{Lei}},
  \bibinfo{journal}{Light: Science \& Applications}
  \textbf{\bibinfo{volume}{9}}, \bibinfo{pages}{1}
  (\bibinfo{year}{2020}{\natexlab{a}}).

\bibitem[{\citenamefont{Zhang et~al.}(2018{\natexlab{a}})\citenamefont{Zhang,
  Ji, Shangguan, Guo, Wang, Huang, Lu, and Zhu}}]{zhang2018strain}
\bibinfo{author}{\bibfnamefont{J.}~\bibnamefont{Zhang}},
  \bibinfo{author}{\bibfnamefont{C.}~\bibnamefont{Ji}},
  \bibinfo{author}{\bibfnamefont{Y.}~\bibnamefont{Shangguan}},
  \bibinfo{author}{\bibfnamefont{B.}~\bibnamefont{Guo}},
  \bibinfo{author}{\bibfnamefont{J.}~\bibnamefont{Wang}},
  \bibinfo{author}{\bibfnamefont{F.}~\bibnamefont{Huang}},
  \bibinfo{author}{\bibfnamefont{X.}~\bibnamefont{Lu}}, \bibnamefont{and}
  \bibinfo{author}{\bibfnamefont{J.}~\bibnamefont{Zhu}},
  \bibinfo{journal}{Physical Review B} \textbf{\bibinfo{volume}{98}},
  \bibinfo{pages}{195133} (\bibinfo{year}{2018}{\natexlab{a}}).

\bibitem[{\citenamefont{Casper et~al.}(2012)\citenamefont{Casper, Graf, Chadov,
  Balke, and Felser}}]{casper2012half}
\bibinfo{author}{\bibfnamefont{F.}~\bibnamefont{Casper}},
  \bibinfo{author}{\bibfnamefont{T.}~\bibnamefont{Graf}},
  \bibinfo{author}{\bibfnamefont{S.}~\bibnamefont{Chadov}},
  \bibinfo{author}{\bibfnamefont{B.}~\bibnamefont{Balke}}, \bibnamefont{and}
  \bibinfo{author}{\bibfnamefont{C.}~\bibnamefont{Felser}},
  \bibinfo{journal}{Semiconductor Science and Technology}
  \textbf{\bibinfo{volume}{27}}, \bibinfo{pages}{063001}
  (\bibinfo{year}{2012}).

\bibitem[{\citenamefont{Bibes and Barthelemy}(2007)}]{bibes2007oxide}
\bibinfo{author}{\bibfnamefont{M.}~\bibnamefont{Bibes}} \bibnamefont{and}
  \bibinfo{author}{\bibfnamefont{A.}~\bibnamefont{Barthelemy}},
  \bibinfo{journal}{IEEE transactions on electron devices}
  \textbf{\bibinfo{volume}{54}}, \bibinfo{pages}{1003} (\bibinfo{year}{2007}).

\bibitem[{\citenamefont{Bibes et~al.}(2011)\citenamefont{Bibes, Villegas, and
  Barth{\'e}l{\'e}my}}]{bibes2011ultrathin}
\bibinfo{author}{\bibfnamefont{M.}~\bibnamefont{Bibes}},
  \bibinfo{author}{\bibfnamefont{J.~E.} \bibnamefont{Villegas}},
  \bibnamefont{and}
  \bibinfo{author}{\bibfnamefont{A.}~\bibnamefont{Barth{\'e}l{\'e}my}},
  \bibinfo{journal}{Advances in Physics} \textbf{\bibinfo{volume}{60}},
  \bibinfo{pages}{5} (\bibinfo{year}{2011}).

\bibitem[{\citenamefont{Eliseev et~al.}(2011)\citenamefont{Eliseev, Glinchuk,
  Khist, Skorokhod, Blinc, and Morozovska}}]{eliseev2011linear}
\bibinfo{author}{\bibfnamefont{E.~A.} \bibnamefont{Eliseev}},
  \bibinfo{author}{\bibfnamefont{M.}~\bibnamefont{Glinchuk}},
  \bibinfo{author}{\bibfnamefont{V.}~\bibnamefont{Khist}},
  \bibinfo{author}{\bibfnamefont{V.}~\bibnamefont{Skorokhod}},
  \bibinfo{author}{\bibfnamefont{R.}~\bibnamefont{Blinc}}, \bibnamefont{and}
  \bibinfo{author}{\bibfnamefont{A.}~\bibnamefont{Morozovska}},
  \bibinfo{journal}{Physical Review B} \textbf{\bibinfo{volume}{84}},
  \bibinfo{pages}{174112} (\bibinfo{year}{2011}).

\bibitem[{\citenamefont{Lukashev and
  Sabirianov}(2010)}]{lukashev2010flexomagnetic}
\bibinfo{author}{\bibfnamefont{P.}~\bibnamefont{Lukashev}} \bibnamefont{and}
  \bibinfo{author}{\bibfnamefont{R.~F.} \bibnamefont{Sabirianov}},
  \bibinfo{journal}{Physical Review B} \textbf{\bibinfo{volume}{82}},
  \bibinfo{pages}{094417} (\bibinfo{year}{2010}).

\bibitem[{\citenamefont{Eliseev et~al.}(2009)\citenamefont{Eliseev, Morozovska,
  Glinchuk, and Blinc}}]{eliseev2009spontaneous}
\bibinfo{author}{\bibfnamefont{E.~A.} \bibnamefont{Eliseev}},
  \bibinfo{author}{\bibfnamefont{A.~N.} \bibnamefont{Morozovska}},
  \bibinfo{author}{\bibfnamefont{M.~D.} \bibnamefont{Glinchuk}},
  \bibnamefont{and} \bibinfo{author}{\bibfnamefont{R.}~\bibnamefont{Blinc}},
  \bibinfo{journal}{Physical Review B} \textbf{\bibinfo{volume}{79}},
  \bibinfo{pages}{165433} (\bibinfo{year}{2009}).

\bibitem[{\citenamefont{Zhang et~al.}(2021)\citenamefont{Zhang, Liu, Dong, Wu,
  Zhang, Wang, Lu, R{\"u}ckriegel, Wang, Duine et~al.}}]{zhang2021strain}
\bibinfo{author}{\bibfnamefont{Y.}~\bibnamefont{Zhang}},
  \bibinfo{author}{\bibfnamefont{J.}~\bibnamefont{Liu}},
  \bibinfo{author}{\bibfnamefont{Y.}~\bibnamefont{Dong}},
  \bibinfo{author}{\bibfnamefont{S.}~\bibnamefont{Wu}},
  \bibinfo{author}{\bibfnamefont{J.}~\bibnamefont{Zhang}},
  \bibinfo{author}{\bibfnamefont{J.}~\bibnamefont{Wang}},
  \bibinfo{author}{\bibfnamefont{J.}~\bibnamefont{Lu}},
  \bibinfo{author}{\bibfnamefont{A.}~\bibnamefont{R{\"u}ckriegel}},
  \bibinfo{author}{\bibfnamefont{H.}~\bibnamefont{Wang}},
  \bibinfo{author}{\bibfnamefont{R.}~\bibnamefont{Duine}},
  \bibnamefont{et~al.}, \bibinfo{journal}{Physical Review Letters}
  \textbf{\bibinfo{volume}{127}}, \bibinfo{pages}{117204}
  (\bibinfo{year}{2021}).

\bibitem[{\citenamefont{Liu et~al.}(2022)\citenamefont{Liu, Chen, and
  Zheng}}]{liu2022flexoresponses}
\bibinfo{author}{\bibfnamefont{L.}~\bibnamefont{Liu}},
  \bibinfo{author}{\bibfnamefont{W.}~\bibnamefont{Chen}}, \bibnamefont{and}
  \bibinfo{author}{\bibfnamefont{Y.}~\bibnamefont{Zheng}},
  \bibinfo{journal}{Physical Review Letters} \textbf{\bibinfo{volume}{128}},
  \bibinfo{pages}{257201} (\bibinfo{year}{2022}).

\bibitem[{\citenamefont{Edstr{\"o}m et~al.}(2022)\citenamefont{Edstr{\"o}m,
  Amoroso, Picozzi, Barone, and Stengel}}]{edstrom2022curved}
\bibinfo{author}{\bibfnamefont{A.}~\bibnamefont{Edstr{\"o}m}},
  \bibinfo{author}{\bibfnamefont{D.}~\bibnamefont{Amoroso}},
  \bibinfo{author}{\bibfnamefont{S.}~\bibnamefont{Picozzi}},
  \bibinfo{author}{\bibfnamefont{P.}~\bibnamefont{Barone}}, \bibnamefont{and}
  \bibinfo{author}{\bibfnamefont{M.}~\bibnamefont{Stengel}},
  \bibinfo{journal}{Physical Review Letters} \textbf{\bibinfo{volume}{128}},
  \bibinfo{pages}{177202} (\bibinfo{year}{2022}).

\bibitem[{\citenamefont{Yanes et~al.}(2019)\citenamefont{Yanes, Garcia-Sanchez,
  Luis, Martinez, Raposo, Torres, and Lopez-Diaz}}]{yanes2019skyrmion}
\bibinfo{author}{\bibfnamefont{R.}~\bibnamefont{Yanes}},
  \bibinfo{author}{\bibfnamefont{F.}~\bibnamefont{Garcia-Sanchez}},
  \bibinfo{author}{\bibfnamefont{R.}~\bibnamefont{Luis}},
  \bibinfo{author}{\bibfnamefont{E.}~\bibnamefont{Martinez}},
  \bibinfo{author}{\bibfnamefont{V.}~\bibnamefont{Raposo}},
  \bibinfo{author}{\bibfnamefont{L.}~\bibnamefont{Torres}}, \bibnamefont{and}
  \bibinfo{author}{\bibfnamefont{L.}~\bibnamefont{Lopez-Diaz}},
  \bibinfo{journal}{Applied Physics Letters} \textbf{\bibinfo{volume}{115}},
  \bibinfo{pages}{132401} (\bibinfo{year}{2019}).

\bibitem[{\citenamefont{Liu et~al.}(2019)\citenamefont{Liu, Huo, Xuan, and
  Yan}}]{liu2019manipulating}
\bibinfo{author}{\bibfnamefont{Y.}~\bibnamefont{Liu}},
  \bibinfo{author}{\bibfnamefont{X.}~\bibnamefont{Huo}},
  \bibinfo{author}{\bibfnamefont{S.}~\bibnamefont{Xuan}}, \bibnamefont{and}
  \bibinfo{author}{\bibfnamefont{H.}~\bibnamefont{Yan}},
  \bibinfo{journal}{Journal of Magnetism and Magnetic Materials}
  \textbf{\bibinfo{volume}{492}}, \bibinfo{pages}{165659}
  (\bibinfo{year}{2019}).

\bibitem[{\citenamefont{Gorshkov et~al.}(2022)\citenamefont{Gorshkov, Gorev,
  Sapozhnikov, and Udalov}}]{gorshkov2022dmi}
\bibinfo{author}{\bibfnamefont{I.~O.} \bibnamefont{Gorshkov}},
  \bibinfo{author}{\bibfnamefont{R.~V.} \bibnamefont{Gorev}},
  \bibinfo{author}{\bibfnamefont{M.~V.} \bibnamefont{Sapozhnikov}},
  \bibnamefont{and} \bibinfo{author}{\bibfnamefont{O.~G.}
  \bibnamefont{Udalov}}, \bibinfo{journal}{ACS Applied Electronic Materials}
  \textbf{\bibinfo{volume}{4}}, \bibinfo{pages}{3205} (\bibinfo{year}{2022}).

\bibitem[{\citenamefont{Xu et~al.}(2020)\citenamefont{Xu, Huang, Barnard, Hong,
  Singh, Wong, Jansen, Harbola, Xiao, Wang et~al.}}]{xu2020strain}
\bibinfo{author}{\bibfnamefont{R.}~\bibnamefont{Xu}},
  \bibinfo{author}{\bibfnamefont{J.}~\bibnamefont{Huang}},
  \bibinfo{author}{\bibfnamefont{E.~S.} \bibnamefont{Barnard}},
  \bibinfo{author}{\bibfnamefont{S.~S.} \bibnamefont{Hong}},
  \bibinfo{author}{\bibfnamefont{P.}~\bibnamefont{Singh}},
  \bibinfo{author}{\bibfnamefont{E.~K.} \bibnamefont{Wong}},
  \bibinfo{author}{\bibfnamefont{T.}~\bibnamefont{Jansen}},
  \bibinfo{author}{\bibfnamefont{V.}~\bibnamefont{Harbola}},
  \bibinfo{author}{\bibfnamefont{J.}~\bibnamefont{Xiao}},
  \bibinfo{author}{\bibfnamefont{B.~Y.} \bibnamefont{Wang}},
  \bibnamefont{et~al.}, \bibinfo{journal}{Nature communications}
  \textbf{\bibinfo{volume}{11}}, \bibinfo{pages}{1} (\bibinfo{year}{2020}).

\bibitem[{\citenamefont{Pesquera et~al.}(2020)\citenamefont{Pesquera,
  Parsonnet, Qualls, Xu, Gubser, Kim, Jiang, Velarde, Huang, Hwang
  et~al.}}]{pesquera2020beyond}
\bibinfo{author}{\bibfnamefont{D.}~\bibnamefont{Pesquera}},
  \bibinfo{author}{\bibfnamefont{E.}~\bibnamefont{Parsonnet}},
  \bibinfo{author}{\bibfnamefont{A.}~\bibnamefont{Qualls}},
  \bibinfo{author}{\bibfnamefont{R.}~\bibnamefont{Xu}},
  \bibinfo{author}{\bibfnamefont{A.~J.} \bibnamefont{Gubser}},
  \bibinfo{author}{\bibfnamefont{J.}~\bibnamefont{Kim}},
  \bibinfo{author}{\bibfnamefont{Y.}~\bibnamefont{Jiang}},
  \bibinfo{author}{\bibfnamefont{G.}~\bibnamefont{Velarde}},
  \bibinfo{author}{\bibfnamefont{Y.-L.} \bibnamefont{Huang}},
  \bibinfo{author}{\bibfnamefont{H.~Y.} \bibnamefont{Hwang}},
  \bibnamefont{et~al.}, \bibinfo{journal}{Advanced Materials}
  \textbf{\bibinfo{volume}{32}}, \bibinfo{pages}{2003780}
  (\bibinfo{year}{2020}).

\bibitem[{\citenamefont{Kogan}(1964)}]{kogan1964piezoelectric}
\bibinfo{author}{\bibfnamefont{S.~M.} \bibnamefont{Kogan}},
  \bibinfo{journal}{Soviet Physics-Solid State} \textbf{\bibinfo{volume}{5}},
  \bibinfo{pages}{2069} (\bibinfo{year}{1964}).

\bibitem[{\citenamefont{Zubko et~al.}(2013)\citenamefont{Zubko, Catalan, and
  Tagantsev}}]{zubko2013flexoelectric}
\bibinfo{author}{\bibfnamefont{P.}~\bibnamefont{Zubko}},
  \bibinfo{author}{\bibfnamefont{G.}~\bibnamefont{Catalan}}, \bibnamefont{and}
  \bibinfo{author}{\bibfnamefont{A.~K.} \bibnamefont{Tagantsev}}
  (\bibinfo{year}{2013}).

\bibitem[{\citenamefont{Cross}(2006)}]{cross2006flexoelectric}
\bibinfo{author}{\bibfnamefont{L.~E.} \bibnamefont{Cross}},
  \bibinfo{journal}{Journal of Materials Science}
  \textbf{\bibinfo{volume}{41}}, \bibinfo{pages}{53} (\bibinfo{year}{2006}).

\bibitem[{\citenamefont{Indenbom et~al.}(1981)\citenamefont{Indenbom, Loginov,
  and Osipov}}]{indenbom1981flexoelectric}
\bibinfo{author}{\bibfnamefont{V.}~\bibnamefont{Indenbom}},
  \bibinfo{author}{\bibfnamefont{E.}~\bibnamefont{Loginov}}, \bibnamefont{and}
  \bibinfo{author}{\bibfnamefont{M.}~\bibnamefont{Osipov}},
  \bibinfo{journal}{Kristallografiya} \textbf{\bibinfo{volume}{26}},
  \bibinfo{pages}{1157} (\bibinfo{year}{1981}).

\bibitem[{\citenamefont{Ma and Cross}(2001)}]{ma2001observation}
\bibinfo{author}{\bibfnamefont{W.}~\bibnamefont{Ma}} \bibnamefont{and}
  \bibinfo{author}{\bibfnamefont{L.~E.} \bibnamefont{Cross}},
  \bibinfo{journal}{Applied Physics Letters} \textbf{\bibinfo{volume}{78}},
  \bibinfo{pages}{2920} (\bibinfo{year}{2001}).

\bibitem[{\citenamefont{Ma and Cross}(2005)}]{ma2005flexoelectric}
\bibinfo{author}{\bibfnamefont{W.}~\bibnamefont{Ma}} \bibnamefont{and}
  \bibinfo{author}{\bibfnamefont{L.~E.} \bibnamefont{Cross}},
  \bibinfo{journal}{Applied Physics Letters} \textbf{\bibinfo{volume}{86}},
  \bibinfo{pages}{072905} (\bibinfo{year}{2005}).

\bibitem[{\citenamefont{Zabalo and Stengel}(2021)}]{zabalo2021switching}
\bibinfo{author}{\bibfnamefont{A.}~\bibnamefont{Zabalo}} \bibnamefont{and}
  \bibinfo{author}{\bibfnamefont{M.}~\bibnamefont{Stengel}},
  \bibinfo{journal}{Physical Review Letters} \textbf{\bibinfo{volume}{126}},
  \bibinfo{pages}{127601} (\bibinfo{year}{2021}).

\bibitem[{\citenamefont{Du et~al.}(2019)\citenamefont{Du, Lim, Zhang,
  Strohbeen, Shourov, Rodolakis, McChesney, Voyles, Fredrickson, and
  Kawasaki}}]{du2019high}
\bibinfo{author}{\bibfnamefont{D.}~\bibnamefont{Du}},
  \bibinfo{author}{\bibfnamefont{A.}~\bibnamefont{Lim}},
  \bibinfo{author}{\bibfnamefont{C.}~\bibnamefont{Zhang}},
  \bibinfo{author}{\bibfnamefont{P.~J.} \bibnamefont{Strohbeen}},
  \bibinfo{author}{\bibfnamefont{E.~H.} \bibnamefont{Shourov}},
  \bibinfo{author}{\bibfnamefont{F.}~\bibnamefont{Rodolakis}},
  \bibinfo{author}{\bibfnamefont{J.~L.} \bibnamefont{McChesney}},
  \bibinfo{author}{\bibfnamefont{P.}~\bibnamefont{Voyles}},
  \bibinfo{author}{\bibfnamefont{D.~C.} \bibnamefont{Fredrickson}},
  \bibnamefont{and} \bibinfo{author}{\bibfnamefont{J.~K.}
  \bibnamefont{Kawasaki}}, \bibinfo{journal}{APL Materials}
  \textbf{\bibinfo{volume}{7}}, \bibinfo{pages}{121107} (\bibinfo{year}{2019}).

\bibitem[{\citenamefont{Peng et~al.}(2020{\natexlab{b}})\citenamefont{Peng,
  Peng, Zhang, Dong, Zhou, Zhou, Li, Liu, Luo, Wang et~al.}}]{peng2020phase}
\bibinfo{author}{\bibfnamefont{B.}~\bibnamefont{Peng}},
  \bibinfo{author}{\bibfnamefont{R.-C.} \bibnamefont{Peng}},
  \bibinfo{author}{\bibfnamefont{Y.-Q.} \bibnamefont{Zhang}},
  \bibinfo{author}{\bibfnamefont{G.}~\bibnamefont{Dong}},
  \bibinfo{author}{\bibfnamefont{Z.}~\bibnamefont{Zhou}},
  \bibinfo{author}{\bibfnamefont{Y.}~\bibnamefont{Zhou}},
  \bibinfo{author}{\bibfnamefont{T.}~\bibnamefont{Li}},
  \bibinfo{author}{\bibfnamefont{Z.}~\bibnamefont{Liu}},
  \bibinfo{author}{\bibfnamefont{Z.}~\bibnamefont{Luo}},
  \bibinfo{author}{\bibfnamefont{S.}~\bibnamefont{Wang}}, \bibnamefont{et~al.},
  \bibinfo{journal}{Science advances} \textbf{\bibinfo{volume}{6}},
  \bibinfo{pages}{eaba5847} (\bibinfo{year}{2020}{\natexlab{b}}).

\bibitem[{\citenamefont{Engelmann et~al.}(2013)\citenamefont{Engelmann,
  Grinenko, Chekhonin, Skrotzki, Efremov, Oswald, Iida, H{\"u}hne, H{\"a}nisch,
  Hoffmann et~al.}}]{engelmann2013strain}
\bibinfo{author}{\bibfnamefont{J.}~\bibnamefont{Engelmann}},
  \bibinfo{author}{\bibfnamefont{V.}~\bibnamefont{Grinenko}},
  \bibinfo{author}{\bibfnamefont{P.}~\bibnamefont{Chekhonin}},
  \bibinfo{author}{\bibfnamefont{W.}~\bibnamefont{Skrotzki}},
  \bibinfo{author}{\bibfnamefont{D.}~\bibnamefont{Efremov}},
  \bibinfo{author}{\bibfnamefont{S.}~\bibnamefont{Oswald}},
  \bibinfo{author}{\bibfnamefont{K.}~\bibnamefont{Iida}},
  \bibinfo{author}{\bibfnamefont{R.}~\bibnamefont{H{\"u}hne}},
  \bibinfo{author}{\bibfnamefont{J.}~\bibnamefont{H{\"a}nisch}},
  \bibinfo{author}{\bibfnamefont{M.}~\bibnamefont{Hoffmann}},
  \bibnamefont{et~al.}, \bibinfo{journal}{Nature communications}
  \textbf{\bibinfo{volume}{4}}, \bibinfo{pages}{1} (\bibinfo{year}{2013}).

\bibitem[{\citenamefont{Tarantini et~al.}(2011)\citenamefont{Tarantini,
  Gurevich, Jaroszynski, Balakirev, Bellingeri, Pallecchi, Ferdeghini, Shen,
  Wen, and Larbalestier}}]{tarantini2011significant}
\bibinfo{author}{\bibfnamefont{C.}~\bibnamefont{Tarantini}},
  \bibinfo{author}{\bibfnamefont{A.}~\bibnamefont{Gurevich}},
  \bibinfo{author}{\bibfnamefont{J.}~\bibnamefont{Jaroszynski}},
  \bibinfo{author}{\bibfnamefont{F.}~\bibnamefont{Balakirev}},
  \bibinfo{author}{\bibfnamefont{E.}~\bibnamefont{Bellingeri}},
  \bibinfo{author}{\bibfnamefont{I.}~\bibnamefont{Pallecchi}},
  \bibinfo{author}{\bibfnamefont{C.}~\bibnamefont{Ferdeghini}},
  \bibinfo{author}{\bibfnamefont{B.}~\bibnamefont{Shen}},
  \bibinfo{author}{\bibfnamefont{H.}~\bibnamefont{Wen}}, \bibnamefont{and}
  \bibinfo{author}{\bibfnamefont{D.}~\bibnamefont{Larbalestier}},
  \bibinfo{journal}{Physical Review B} \textbf{\bibinfo{volume}{84}},
  \bibinfo{pages}{184522} (\bibinfo{year}{2011}).

\bibitem[{\citenamefont{Okabe et~al.}(2010)\citenamefont{Okabe, Takeshita,
  Horigane, Muranaka, and Akimitsu}}]{okabe2010pressure}
\bibinfo{author}{\bibfnamefont{H.}~\bibnamefont{Okabe}},
  \bibinfo{author}{\bibfnamefont{N.}~\bibnamefont{Takeshita}},
  \bibinfo{author}{\bibfnamefont{K.}~\bibnamefont{Horigane}},
  \bibinfo{author}{\bibfnamefont{T.}~\bibnamefont{Muranaka}}, \bibnamefont{and}
  \bibinfo{author}{\bibfnamefont{J.}~\bibnamefont{Akimitsu}},
  \bibinfo{journal}{Physical Review B} \textbf{\bibinfo{volume}{81}},
  \bibinfo{pages}{205119} (\bibinfo{year}{2010}).

\bibitem[{\citenamefont{Mandal et~al.}(2017)\citenamefont{Mandal, Zhang,
  Ismail-Beigi, and Haule}}]{mandal2017correlated}
\bibinfo{author}{\bibfnamefont{S.}~\bibnamefont{Mandal}},
  \bibinfo{author}{\bibfnamefont{P.}~\bibnamefont{Zhang}},
  \bibinfo{author}{\bibfnamefont{S.}~\bibnamefont{Ismail-Beigi}},
  \bibnamefont{and} \bibinfo{author}{\bibfnamefont{K.}~\bibnamefont{Haule}},
  \bibinfo{journal}{Physical review letters} \textbf{\bibinfo{volume}{119}},
  \bibinfo{pages}{067004} (\bibinfo{year}{2017}).

\bibitem[{\citenamefont{Wang et~al.}(2012)\citenamefont{Wang, Li, Zhang, Zhang,
  Zhang, Li, Ding, Ou, Deng, Chang et~al.}}]{wang2012interface}
\bibinfo{author}{\bibfnamefont{Q.-Y.} \bibnamefont{Wang}},
  \bibinfo{author}{\bibfnamefont{Z.}~\bibnamefont{Li}},
  \bibinfo{author}{\bibfnamefont{W.-H.} \bibnamefont{Zhang}},
  \bibinfo{author}{\bibfnamefont{Z.-C.} \bibnamefont{Zhang}},
  \bibinfo{author}{\bibfnamefont{J.-S.} \bibnamefont{Zhang}},
  \bibinfo{author}{\bibfnamefont{W.}~\bibnamefont{Li}},
  \bibinfo{author}{\bibfnamefont{H.}~\bibnamefont{Ding}},
  \bibinfo{author}{\bibfnamefont{Y.-B.} \bibnamefont{Ou}},
  \bibinfo{author}{\bibfnamefont{P.}~\bibnamefont{Deng}},
  \bibinfo{author}{\bibfnamefont{K.}~\bibnamefont{Chang}},
  \bibnamefont{et~al.}, \bibinfo{journal}{Chinese Physics Letters}
  \textbf{\bibinfo{volume}{29}}, \bibinfo{pages}{037402}
  (\bibinfo{year}{2012}).

\bibitem[{\citenamefont{Lee et~al.}(2014)\citenamefont{Lee, Schmitt, Moore,
  Johnston, Cui, Li, Yi, Liu, Hashimoto, Zhang et~al.}}]{lee2014interfacial}
\bibinfo{author}{\bibfnamefont{J.}~\bibnamefont{Lee}},
  \bibinfo{author}{\bibfnamefont{F.}~\bibnamefont{Schmitt}},
  \bibinfo{author}{\bibfnamefont{R.}~\bibnamefont{Moore}},
  \bibinfo{author}{\bibfnamefont{S.}~\bibnamefont{Johnston}},
  \bibinfo{author}{\bibfnamefont{Y.-T.} \bibnamefont{Cui}},
  \bibinfo{author}{\bibfnamefont{W.}~\bibnamefont{Li}},
  \bibinfo{author}{\bibfnamefont{M.}~\bibnamefont{Yi}},
  \bibinfo{author}{\bibfnamefont{Z.}~\bibnamefont{Liu}},
  \bibinfo{author}{\bibfnamefont{M.}~\bibnamefont{Hashimoto}},
  \bibinfo{author}{\bibfnamefont{Y.}~\bibnamefont{Zhang}},
  \bibnamefont{et~al.}, \bibinfo{journal}{Nature}
  \textbf{\bibinfo{volume}{515}}, \bibinfo{pages}{245} (\bibinfo{year}{2014}).

\bibitem[{\citenamefont{Tan et~al.}(2013)\citenamefont{Tan, Zhang, Xia, Ye,
  Chen, Xie, Peng, Xu, Fan, Xu et~al.}}]{tan2013interface}
\bibinfo{author}{\bibfnamefont{S.}~\bibnamefont{Tan}},
  \bibinfo{author}{\bibfnamefont{Y.}~\bibnamefont{Zhang}},
  \bibinfo{author}{\bibfnamefont{M.}~\bibnamefont{Xia}},
  \bibinfo{author}{\bibfnamefont{Z.}~\bibnamefont{Ye}},
  \bibinfo{author}{\bibfnamefont{F.}~\bibnamefont{Chen}},
  \bibinfo{author}{\bibfnamefont{X.}~\bibnamefont{Xie}},
  \bibinfo{author}{\bibfnamefont{R.}~\bibnamefont{Peng}},
  \bibinfo{author}{\bibfnamefont{D.}~\bibnamefont{Xu}},
  \bibinfo{author}{\bibfnamefont{Q.}~\bibnamefont{Fan}},
  \bibinfo{author}{\bibfnamefont{H.}~\bibnamefont{Xu}}, \bibnamefont{et~al.},
  \bibinfo{journal}{Nature materials} \textbf{\bibinfo{volume}{12}},
  \bibinfo{pages}{634} (\bibinfo{year}{2013}).

\bibitem[{\citenamefont{Locquet et~al.}(1998)\citenamefont{Locquet, Perret,
  Fompeyrine, M{\"a}chler, Seo, and Van~Tendeloo}}]{locquet1998doubling}
\bibinfo{author}{\bibfnamefont{J.-P.} \bibnamefont{Locquet}},
  \bibinfo{author}{\bibfnamefont{J.}~\bibnamefont{Perret}},
  \bibinfo{author}{\bibfnamefont{J.}~\bibnamefont{Fompeyrine}},
  \bibinfo{author}{\bibfnamefont{E.}~\bibnamefont{M{\"a}chler}},
  \bibinfo{author}{\bibfnamefont{J.~W.} \bibnamefont{Seo}}, \bibnamefont{and}
  \bibinfo{author}{\bibfnamefont{G.}~\bibnamefont{Van~Tendeloo}},
  \bibinfo{journal}{Nature} \textbf{\bibinfo{volume}{394}},
  \bibinfo{pages}{453} (\bibinfo{year}{1998}).

\bibitem[{\citenamefont{Bozovic et~al.}(2002)\citenamefont{Bozovic, Logvenov,
  Belca, Narimbetov, and Sveklo}}]{bozovic2002epitaxial}
\bibinfo{author}{\bibfnamefont{I.}~\bibnamefont{Bozovic}},
  \bibinfo{author}{\bibfnamefont{G.}~\bibnamefont{Logvenov}},
  \bibinfo{author}{\bibfnamefont{I.}~\bibnamefont{Belca}},
  \bibinfo{author}{\bibfnamefont{B.}~\bibnamefont{Narimbetov}},
  \bibnamefont{and} \bibinfo{author}{\bibfnamefont{I.}~\bibnamefont{Sveklo}},
  \bibinfo{journal}{Physical review letters} \textbf{\bibinfo{volume}{89}},
  \bibinfo{pages}{107001} (\bibinfo{year}{2002}).

\bibitem[{\citenamefont{Edge et~al.}(2015)\citenamefont{Edge, Kedem, Aschauer,
  Spaldin, and Balatsky}}]{edge2015quantum}
\bibinfo{author}{\bibfnamefont{J.~M.} \bibnamefont{Edge}},
  \bibinfo{author}{\bibfnamefont{Y.}~\bibnamefont{Kedem}},
  \bibinfo{author}{\bibfnamefont{U.}~\bibnamefont{Aschauer}},
  \bibinfo{author}{\bibfnamefont{N.~A.} \bibnamefont{Spaldin}},
  \bibnamefont{and} \bibinfo{author}{\bibfnamefont{A.~V.}
  \bibnamefont{Balatsky}}, \bibinfo{journal}{Physical Review Letters}
  \textbf{\bibinfo{volume}{115}}, \bibinfo{pages}{247002}
  (\bibinfo{year}{2015}).

\bibitem[{\citenamefont{Ahadi et~al.}(2019)\citenamefont{Ahadi, Galletti, Li,
  Salmani-Rezaie, Wu, and Stemmer}}]{ahadi2019enhancing}
\bibinfo{author}{\bibfnamefont{K.}~\bibnamefont{Ahadi}},
  \bibinfo{author}{\bibfnamefont{L.}~\bibnamefont{Galletti}},
  \bibinfo{author}{\bibfnamefont{Y.}~\bibnamefont{Li}},
  \bibinfo{author}{\bibfnamefont{S.}~\bibnamefont{Salmani-Rezaie}},
  \bibinfo{author}{\bibfnamefont{W.}~\bibnamefont{Wu}}, \bibnamefont{and}
  \bibinfo{author}{\bibfnamefont{S.}~\bibnamefont{Stemmer}},
  \bibinfo{journal}{Science advances} \textbf{\bibinfo{volume}{5}},
  \bibinfo{pages}{eaaw0120} (\bibinfo{year}{2019}).

\bibitem[{\citenamefont{M{\"u}ller and Burkard}(1979)}]{muller1979srti}
\bibinfo{author}{\bibfnamefont{K.~A.} \bibnamefont{M{\"u}ller}}
  \bibnamefont{and} \bibinfo{author}{\bibfnamefont{H.}~\bibnamefont{Burkard}},
  \bibinfo{journal}{Physical Review B} \textbf{\bibinfo{volume}{19}},
  \bibinfo{pages}{3593} (\bibinfo{year}{1979}).

\bibitem[{\citenamefont{Rytz et~al.}(1980)\citenamefont{Rytz, H{\"o}chli, and
  Bilz}}]{rytz1980dielectric}
\bibinfo{author}{\bibfnamefont{D.}~\bibnamefont{Rytz}},
  \bibinfo{author}{\bibfnamefont{U.}~\bibnamefont{H{\"o}chli}},
  \bibnamefont{and} \bibinfo{author}{\bibfnamefont{H.}~\bibnamefont{Bilz}},
  \bibinfo{journal}{Physical Review B} \textbf{\bibinfo{volume}{22}},
  \bibinfo{pages}{359} (\bibinfo{year}{1980}).

\bibitem[{\citenamefont{Yip}(2014)}]{yip2014noncentrosymmetric}
\bibinfo{author}{\bibfnamefont{S.}~\bibnamefont{Yip}}, \bibinfo{journal}{Annu.
  Rev. Condens. Matter Phys.} \textbf{\bibinfo{volume}{5}}, \bibinfo{pages}{15}
  (\bibinfo{year}{2014}).

\bibitem[{\citenamefont{Gor'kov and Rashba}(2001)}]{gor2001superconducting}
\bibinfo{author}{\bibfnamefont{L.~P.} \bibnamefont{Gor'kov}} \bibnamefont{and}
  \bibinfo{author}{\bibfnamefont{E.~I.} \bibnamefont{Rashba}},
  \bibinfo{journal}{Physical Review Letters} \textbf{\bibinfo{volume}{87}},
  \bibinfo{pages}{037004} (\bibinfo{year}{2001}).

\bibitem[{\citenamefont{Zhang et~al.}(2018{\natexlab{b}})\citenamefont{Zhang,
  Yaji, Hashimoto, Ota, Kondo, Okazaki, Wang, Wen, Gu, Ding
  et~al.}}]{zhang2018observation}
\bibinfo{author}{\bibfnamefont{P.}~\bibnamefont{Zhang}},
  \bibinfo{author}{\bibfnamefont{K.}~\bibnamefont{Yaji}},
  \bibinfo{author}{\bibfnamefont{T.}~\bibnamefont{Hashimoto}},
  \bibinfo{author}{\bibfnamefont{Y.}~\bibnamefont{Ota}},
  \bibinfo{author}{\bibfnamefont{T.}~\bibnamefont{Kondo}},
  \bibinfo{author}{\bibfnamefont{K.}~\bibnamefont{Okazaki}},
  \bibinfo{author}{\bibfnamefont{Z.}~\bibnamefont{Wang}},
  \bibinfo{author}{\bibfnamefont{J.}~\bibnamefont{Wen}},
  \bibinfo{author}{\bibfnamefont{G.}~\bibnamefont{Gu}},
  \bibinfo{author}{\bibfnamefont{H.}~\bibnamefont{Ding}}, \bibnamefont{et~al.},
  \bibinfo{journal}{Science} \textbf{\bibinfo{volume}{360}},
  \bibinfo{pages}{182} (\bibinfo{year}{2018}{\natexlab{b}}).

\bibitem[{\citenamefont{Wang et~al.}(2018)\citenamefont{Wang, Kong, Fan, Chen,
  Zhu, Liu, Cao, Sun, Du, Schneeloch et~al.}}]{wang2018evidence}
\bibinfo{author}{\bibfnamefont{D.}~\bibnamefont{Wang}},
  \bibinfo{author}{\bibfnamefont{L.}~\bibnamefont{Kong}},
  \bibinfo{author}{\bibfnamefont{P.}~\bibnamefont{Fan}},
  \bibinfo{author}{\bibfnamefont{H.}~\bibnamefont{Chen}},
  \bibinfo{author}{\bibfnamefont{S.}~\bibnamefont{Zhu}},
  \bibinfo{author}{\bibfnamefont{W.}~\bibnamefont{Liu}},
  \bibinfo{author}{\bibfnamefont{L.}~\bibnamefont{Cao}},
  \bibinfo{author}{\bibfnamefont{Y.}~\bibnamefont{Sun}},
  \bibinfo{author}{\bibfnamefont{S.}~\bibnamefont{Du}},
  \bibinfo{author}{\bibfnamefont{J.}~\bibnamefont{Schneeloch}},
  \bibnamefont{et~al.}, \bibinfo{journal}{Science}
  \textbf{\bibinfo{volume}{362}}, \bibinfo{pages}{333} (\bibinfo{year}{2018}).

\bibitem[{\citenamefont{Machida et~al.}(2019)\citenamefont{Machida, Sun, Pyon,
  Takeda, Kohsaka, Hanaguri, Sasagawa, and Tamegai}}]{machida2019zero}
\bibinfo{author}{\bibfnamefont{T.}~\bibnamefont{Machida}},
  \bibinfo{author}{\bibfnamefont{Y.}~\bibnamefont{Sun}},
  \bibinfo{author}{\bibfnamefont{S.}~\bibnamefont{Pyon}},
  \bibinfo{author}{\bibfnamefont{S.}~\bibnamefont{Takeda}},
  \bibinfo{author}{\bibfnamefont{Y.}~\bibnamefont{Kohsaka}},
  \bibinfo{author}{\bibfnamefont{T.}~\bibnamefont{Hanaguri}},
  \bibinfo{author}{\bibfnamefont{T.}~\bibnamefont{Sasagawa}}, \bibnamefont{and}
  \bibinfo{author}{\bibfnamefont{T.}~\bibnamefont{Tamegai}},
  \bibinfo{journal}{Nature materials} \textbf{\bibinfo{volume}{18}},
  \bibinfo{pages}{811} (\bibinfo{year}{2019}).

\bibitem[{\citenamefont{Lohani et~al.}(2020)\citenamefont{Lohani, Hazra, Ribak,
  Nitzav, Fu, Yan, Randeria, and Kanigel}}]{lohani2020band}
\bibinfo{author}{\bibfnamefont{H.}~\bibnamefont{Lohani}},
  \bibinfo{author}{\bibfnamefont{T.}~\bibnamefont{Hazra}},
  \bibinfo{author}{\bibfnamefont{A.}~\bibnamefont{Ribak}},
  \bibinfo{author}{\bibfnamefont{Y.}~\bibnamefont{Nitzav}},
  \bibinfo{author}{\bibfnamefont{H.}~\bibnamefont{Fu}},
  \bibinfo{author}{\bibfnamefont{B.}~\bibnamefont{Yan}},
  \bibinfo{author}{\bibfnamefont{M.}~\bibnamefont{Randeria}}, \bibnamefont{and}
  \bibinfo{author}{\bibfnamefont{A.}~\bibnamefont{Kanigel}},
  \bibinfo{journal}{Physical Review B} \textbf{\bibinfo{volume}{101}},
  \bibinfo{pages}{245146} (\bibinfo{year}{2020}).

\bibitem[{\citenamefont{Wang et~al.}(2015)\citenamefont{Wang, Zhang, Xu, Zeng,
  Miao, Xu, Qian, Weng, Richard, Fedorov et~al.}}]{wang2015topological}
\bibinfo{author}{\bibfnamefont{Z.}~\bibnamefont{Wang}},
  \bibinfo{author}{\bibfnamefont{P.}~\bibnamefont{Zhang}},
  \bibinfo{author}{\bibfnamefont{G.}~\bibnamefont{Xu}},
  \bibinfo{author}{\bibfnamefont{L.}~\bibnamefont{Zeng}},
  \bibinfo{author}{\bibfnamefont{H.}~\bibnamefont{Miao}},
  \bibinfo{author}{\bibfnamefont{X.}~\bibnamefont{Xu}},
  \bibinfo{author}{\bibfnamefont{T.}~\bibnamefont{Qian}},
  \bibinfo{author}{\bibfnamefont{H.}~\bibnamefont{Weng}},
  \bibinfo{author}{\bibfnamefont{P.}~\bibnamefont{Richard}},
  \bibinfo{author}{\bibfnamefont{A.~V.} \bibnamefont{Fedorov}},
  \bibnamefont{et~al.}, \bibinfo{journal}{Physical Review B}
  \textbf{\bibinfo{volume}{92}}, \bibinfo{pages}{115119}
  (\bibinfo{year}{2015}).

\bibitem[{\citenamefont{Lin et~al.}(2010)\citenamefont{Lin, Wray, Xia, Xu, Jia,
  Cava, Bansil, and Hasan}}]{lin2010half}
\bibinfo{author}{\bibfnamefont{H.}~\bibnamefont{Lin}},
  \bibinfo{author}{\bibfnamefont{L.~A.} \bibnamefont{Wray}},
  \bibinfo{author}{\bibfnamefont{Y.}~\bibnamefont{Xia}},
  \bibinfo{author}{\bibfnamefont{S.}~\bibnamefont{Xu}},
  \bibinfo{author}{\bibfnamefont{S.}~\bibnamefont{Jia}},
  \bibinfo{author}{\bibfnamefont{R.~J.} \bibnamefont{Cava}},
  \bibinfo{author}{\bibfnamefont{A.}~\bibnamefont{Bansil}}, \bibnamefont{and}
  \bibinfo{author}{\bibfnamefont{M.~Z.} \bibnamefont{Hasan}},
  \bibinfo{journal}{Nature materials} \textbf{\bibinfo{volume}{9}},
  \bibinfo{pages}{546} (\bibinfo{year}{2010}).

\bibitem[{\citenamefont{Chadov et~al.}(2010)\citenamefont{Chadov, Qi,
  K{\"u}bler, Fecher, Felser, and Zhang}}]{chadov2010tunable}
\bibinfo{author}{\bibfnamefont{S.}~\bibnamefont{Chadov}},
  \bibinfo{author}{\bibfnamefont{X.}~\bibnamefont{Qi}},
  \bibinfo{author}{\bibfnamefont{J.}~\bibnamefont{K{\"u}bler}},
  \bibinfo{author}{\bibfnamefont{G.~H.} \bibnamefont{Fecher}},
  \bibinfo{author}{\bibfnamefont{C.}~\bibnamefont{Felser}}, \bibnamefont{and}
  \bibinfo{author}{\bibfnamefont{S.~C.} \bibnamefont{Zhang}},
  \bibinfo{journal}{Nature materials} \textbf{\bibinfo{volume}{9}},
  \bibinfo{pages}{541} (\bibinfo{year}{2010}).

\bibitem[{\citenamefont{Nakajima et~al.}(2015)\citenamefont{Nakajima, Hu,
  Kirshenbaum, Hughes, Syers, Wang, Wang, Wang, Saha, Pratt
  et~al.}}]{nakajima2015topological}
\bibinfo{author}{\bibfnamefont{Y.}~\bibnamefont{Nakajima}},
  \bibinfo{author}{\bibfnamefont{R.}~\bibnamefont{Hu}},
  \bibinfo{author}{\bibfnamefont{K.}~\bibnamefont{Kirshenbaum}},
  \bibinfo{author}{\bibfnamefont{A.}~\bibnamefont{Hughes}},
  \bibinfo{author}{\bibfnamefont{P.}~\bibnamefont{Syers}},
  \bibinfo{author}{\bibfnamefont{X.}~\bibnamefont{Wang}},
  \bibinfo{author}{\bibfnamefont{K.}~\bibnamefont{Wang}},
  \bibinfo{author}{\bibfnamefont{R.}~\bibnamefont{Wang}},
  \bibinfo{author}{\bibfnamefont{S.~R.} \bibnamefont{Saha}},
  \bibinfo{author}{\bibfnamefont{D.}~\bibnamefont{Pratt}},
  \bibnamefont{et~al.}, \bibinfo{journal}{Science advances}
  \textbf{\bibinfo{volume}{1}}, \bibinfo{pages}{e1500242}
  (\bibinfo{year}{2015}).

\bibitem[{\citenamefont{Singh et~al.}(2020)\citenamefont{Singh, Ramarao, and
  Peter}}]{singh2020rare}
\bibinfo{author}{\bibfnamefont{A.~K.} \bibnamefont{Singh}},
  \bibinfo{author}{\bibfnamefont{S.}~\bibnamefont{Ramarao}}, \bibnamefont{and}
  \bibinfo{author}{\bibfnamefont{S.~C.} \bibnamefont{Peter}},
  \bibinfo{journal}{APL Materials} \textbf{\bibinfo{volume}{8}},
  \bibinfo{pages}{060903} (\bibinfo{year}{2020}).

\bibitem[{\citenamefont{Al-Sawai et~al.}(2010)\citenamefont{Al-Sawai, Lin,
  Markiewicz, Wray, Xia, Xu, Hasan, and Bansil}}]{al2010topological}
\bibinfo{author}{\bibfnamefont{W.}~\bibnamefont{Al-Sawai}},
  \bibinfo{author}{\bibfnamefont{H.}~\bibnamefont{Lin}},
  \bibinfo{author}{\bibfnamefont{R.}~\bibnamefont{Markiewicz}},
  \bibinfo{author}{\bibfnamefont{L.}~\bibnamefont{Wray}},
  \bibinfo{author}{\bibfnamefont{Y.}~\bibnamefont{Xia}},
  \bibinfo{author}{\bibfnamefont{S.-Y.} \bibnamefont{Xu}},
  \bibinfo{author}{\bibfnamefont{M.}~\bibnamefont{Hasan}}, \bibnamefont{and}
  \bibinfo{author}{\bibfnamefont{A.}~\bibnamefont{Bansil}},
  \bibinfo{journal}{Physical Review B} \textbf{\bibinfo{volume}{82}},
  \bibinfo{pages}{125208} (\bibinfo{year}{2010}).

\bibitem[{\citenamefont{Xiao et~al.}(2010)\citenamefont{Xiao, Yao, Feng, Wen,
  Zhu, Chen, Stocks, and Zhang}}]{xiao2010half}
\bibinfo{author}{\bibfnamefont{D.}~\bibnamefont{Xiao}},
  \bibinfo{author}{\bibfnamefont{Y.}~\bibnamefont{Yao}},
  \bibinfo{author}{\bibfnamefont{W.}~\bibnamefont{Feng}},
  \bibinfo{author}{\bibfnamefont{J.}~\bibnamefont{Wen}},
  \bibinfo{author}{\bibfnamefont{W.}~\bibnamefont{Zhu}},
  \bibinfo{author}{\bibfnamefont{X.-Q.} \bibnamefont{Chen}},
  \bibinfo{author}{\bibfnamefont{G.~M.} \bibnamefont{Stocks}},
  \bibnamefont{and} \bibinfo{author}{\bibfnamefont{Z.}~\bibnamefont{Zhang}},
  \bibinfo{journal}{Physical review letters} \textbf{\bibinfo{volume}{105}},
  \bibinfo{pages}{096404} (\bibinfo{year}{2010}).

\bibitem[{\citenamefont{Jiang et~al.}(2017)\citenamefont{Jiang, Mao, Duan, Lai,
  Watanabe, Taniguchi, and Andrei}}]{jiang2017visualizing}
\bibinfo{author}{\bibfnamefont{Y.}~\bibnamefont{Jiang}},
  \bibinfo{author}{\bibfnamefont{J.}~\bibnamefont{Mao}},
  \bibinfo{author}{\bibfnamefont{J.}~\bibnamefont{Duan}},
  \bibinfo{author}{\bibfnamefont{X.}~\bibnamefont{Lai}},
  \bibinfo{author}{\bibfnamefont{K.}~\bibnamefont{Watanabe}},
  \bibinfo{author}{\bibfnamefont{T.}~\bibnamefont{Taniguchi}},
  \bibnamefont{and} \bibinfo{author}{\bibfnamefont{E.~Y.}
  \bibnamefont{Andrei}}, \bibinfo{journal}{Nano letters}
  \textbf{\bibinfo{volume}{17}}, \bibinfo{pages}{2839} (\bibinfo{year}{2017}).

\bibitem[{\citenamefont{Kang et~al.}(2021)\citenamefont{Kang, Sun, Luo, Lu,
  Chen, Kim, Jung, Gao, Parluhutan, Ge et~al.}}]{kang2021pseudo}
\bibinfo{author}{\bibfnamefont{D.-H.} \bibnamefont{Kang}},
  \bibinfo{author}{\bibfnamefont{H.}~\bibnamefont{Sun}},
  \bibinfo{author}{\bibfnamefont{M.}~\bibnamefont{Luo}},
  \bibinfo{author}{\bibfnamefont{K.}~\bibnamefont{Lu}},
  \bibinfo{author}{\bibfnamefont{M.}~\bibnamefont{Chen}},
  \bibinfo{author}{\bibfnamefont{Y.}~\bibnamefont{Kim}},
  \bibinfo{author}{\bibfnamefont{Y.}~\bibnamefont{Jung}},
  \bibinfo{author}{\bibfnamefont{X.}~\bibnamefont{Gao}},
  \bibinfo{author}{\bibfnamefont{S.~J.} \bibnamefont{Parluhutan}},
  \bibinfo{author}{\bibfnamefont{J.}~\bibnamefont{Ge}}, \bibnamefont{et~al.},
  \bibinfo{journal}{Nature Communications} \textbf{\bibinfo{volume}{12}},
  \bibinfo{pages}{1} (\bibinfo{year}{2021}).

\bibitem[{\citenamefont{Tang and Fu}(2014)}]{tang2014strain}
\bibinfo{author}{\bibfnamefont{E.}~\bibnamefont{Tang}} \bibnamefont{and}
  \bibinfo{author}{\bibfnamefont{L.}~\bibnamefont{Fu}},
  \bibinfo{journal}{Nature Physics} \textbf{\bibinfo{volume}{10}},
  \bibinfo{pages}{964} (\bibinfo{year}{2014}).

\bibitem[{\citenamefont{Armitage et~al.}(2018)\citenamefont{Armitage, Mele, and
  Vishwanath}}]{armitage2018weyl}
\bibinfo{author}{\bibfnamefont{N.}~\bibnamefont{Armitage}},
  \bibinfo{author}{\bibfnamefont{E.}~\bibnamefont{Mele}}, \bibnamefont{and}
  \bibinfo{author}{\bibfnamefont{A.}~\bibnamefont{Vishwanath}},
  \bibinfo{journal}{Reviews of Modern Physics} \textbf{\bibinfo{volume}{90}},
  \bibinfo{pages}{015001} (\bibinfo{year}{2018}).

\bibitem[{\citenamefont{Roberts et~al.}(2006)\citenamefont{Roberts, Klein,
  Savage, Slinker, Friesen, Celler, Eriksson, and
  Lagally}}]{roberts2006elastically}
\bibinfo{author}{\bibfnamefont{M.~M.} \bibnamefont{Roberts}},
  \bibinfo{author}{\bibfnamefont{L.~J.} \bibnamefont{Klein}},
  \bibinfo{author}{\bibfnamefont{D.~E.} \bibnamefont{Savage}},
  \bibinfo{author}{\bibfnamefont{K.~A.} \bibnamefont{Slinker}},
  \bibinfo{author}{\bibfnamefont{M.}~\bibnamefont{Friesen}},
  \bibinfo{author}{\bibfnamefont{G.}~\bibnamefont{Celler}},
  \bibinfo{author}{\bibfnamefont{M.~A.} \bibnamefont{Eriksson}},
  \bibnamefont{and} \bibinfo{author}{\bibfnamefont{M.~G.}
  \bibnamefont{Lagally}}, \bibinfo{journal}{Nature materials}
  \textbf{\bibinfo{volume}{5}}, \bibinfo{pages}{388} (\bibinfo{year}{2006}).

\bibitem[{\citenamefont{Cheng et~al.}(2013)\citenamefont{Cheng, Shiu, Li, Han,
  Shi, and Sadana}}]{cheng2013epitaxial}
\bibinfo{author}{\bibfnamefont{C.-W.} \bibnamefont{Cheng}},
  \bibinfo{author}{\bibfnamefont{K.-T.} \bibnamefont{Shiu}},
  \bibinfo{author}{\bibfnamefont{N.}~\bibnamefont{Li}},
  \bibinfo{author}{\bibfnamefont{S.-J.} \bibnamefont{Han}},
  \bibinfo{author}{\bibfnamefont{L.}~\bibnamefont{Shi}}, \bibnamefont{and}
  \bibinfo{author}{\bibfnamefont{D.~K.} \bibnamefont{Sadana}},
  \bibinfo{journal}{Nature communications} \textbf{\bibinfo{volume}{4}},
  \bibinfo{pages}{1} (\bibinfo{year}{2013}).

\bibitem[{\citenamefont{Dong et~al.}(2004)\citenamefont{Dong, Xie, Lu,
  Adelmann, Palmstr{\o}m, Cui, Pan, Shield, James, and
  McKernan}}]{dong2004shape}
\bibinfo{author}{\bibfnamefont{J.}~\bibnamefont{Dong}},
  \bibinfo{author}{\bibfnamefont{J.}~\bibnamefont{Xie}},
  \bibinfo{author}{\bibfnamefont{J.}~\bibnamefont{Lu}},
  \bibinfo{author}{\bibfnamefont{C.}~\bibnamefont{Adelmann}},
  \bibinfo{author}{\bibfnamefont{C.}~\bibnamefont{Palmstr{\o}m}},
  \bibinfo{author}{\bibfnamefont{J.}~\bibnamefont{Cui}},
  \bibinfo{author}{\bibfnamefont{Q.}~\bibnamefont{Pan}},
  \bibinfo{author}{\bibfnamefont{T.}~\bibnamefont{Shield}},
  \bibinfo{author}{\bibfnamefont{R.}~\bibnamefont{James}}, \bibnamefont{and}
  \bibinfo{author}{\bibfnamefont{S.}~\bibnamefont{McKernan}},
  \bibinfo{journal}{Journal of applied physics} \textbf{\bibinfo{volume}{95}},
  \bibinfo{pages}{2593} (\bibinfo{year}{2004}).

\bibitem[{\citenamefont{Ji et~al.}(2019)\citenamefont{Ji, Cai, Paudel, Sun,
  Zhang, Han, Wei, Zang, Gu, Zhang et~al.}}]{ji2019freestanding}
\bibinfo{author}{\bibfnamefont{D.}~\bibnamefont{Ji}},
  \bibinfo{author}{\bibfnamefont{S.}~\bibnamefont{Cai}},
  \bibinfo{author}{\bibfnamefont{T.~R.} \bibnamefont{Paudel}},
  \bibinfo{author}{\bibfnamefont{H.}~\bibnamefont{Sun}},
  \bibinfo{author}{\bibfnamefont{C.}~\bibnamefont{Zhang}},
  \bibinfo{author}{\bibfnamefont{L.}~\bibnamefont{Han}},
  \bibinfo{author}{\bibfnamefont{Y.}~\bibnamefont{Wei}},
  \bibinfo{author}{\bibfnamefont{Y.}~\bibnamefont{Zang}},
  \bibinfo{author}{\bibfnamefont{M.}~\bibnamefont{Gu}},
  \bibinfo{author}{\bibfnamefont{Y.}~\bibnamefont{Zhang}},
  \bibnamefont{et~al.}, \bibinfo{journal}{Nature}
  \textbf{\bibinfo{volume}{570}}, \bibinfo{pages}{87} (\bibinfo{year}{2019}).

\bibitem[{\citenamefont{Chiabrera et~al.}(2022)\citenamefont{Chiabrera, Yun,
  Li, Dahm, Zhang, Kirchert, Christensen, Trier, Jespersen, and
  Pryds}}]{chiabrera2022freestanding}
\bibinfo{author}{\bibfnamefont{F.~M.} \bibnamefont{Chiabrera}},
  \bibinfo{author}{\bibfnamefont{S.}~\bibnamefont{Yun}},
  \bibinfo{author}{\bibfnamefont{Y.}~\bibnamefont{Li}},
  \bibinfo{author}{\bibfnamefont{R.~T.} \bibnamefont{Dahm}},
  \bibinfo{author}{\bibfnamefont{H.}~\bibnamefont{Zhang}},
  \bibinfo{author}{\bibfnamefont{C.~K.} \bibnamefont{Kirchert}},
  \bibinfo{author}{\bibfnamefont{D.~V.} \bibnamefont{Christensen}},
  \bibinfo{author}{\bibfnamefont{F.}~\bibnamefont{Trier}},
  \bibinfo{author}{\bibfnamefont{T.~S.} \bibnamefont{Jespersen}},
  \bibnamefont{and} \bibinfo{author}{\bibfnamefont{N.}~\bibnamefont{Pryds}},
  \bibinfo{journal}{Annalen der Physik} \textbf{\bibinfo{volume}{534}},
  \bibinfo{pages}{2200084} (\bibinfo{year}{2022}).

\bibitem[{\citenamefont{Bourlier et~al.}(2020)\citenamefont{Bourlier, Berini,
  Fr{\'e}gnaux, Fouchet, Aureau, and Dumont}}]{bourlier2020transfer}
\bibinfo{author}{\bibfnamefont{Y.}~\bibnamefont{Bourlier}},
  \bibinfo{author}{\bibfnamefont{B.}~\bibnamefont{Berini}},
  \bibinfo{author}{\bibfnamefont{M.}~\bibnamefont{Fr{\'e}gnaux}},
  \bibinfo{author}{\bibfnamefont{A.}~\bibnamefont{Fouchet}},
  \bibinfo{author}{\bibfnamefont{D.}~\bibnamefont{Aureau}}, \bibnamefont{and}
  \bibinfo{author}{\bibfnamefont{Y.}~\bibnamefont{Dumont}},
  \bibinfo{journal}{ACS applied materials \& interfaces}
  \textbf{\bibinfo{volume}{12}}, \bibinfo{pages}{8466} (\bibinfo{year}{2020}).

\bibitem[{\citenamefont{Takahashi and
  Lippmaa}(2020)}]{takahashi2020sacrificial}
\bibinfo{author}{\bibfnamefont{R.}~\bibnamefont{Takahashi}} \bibnamefont{and}
  \bibinfo{author}{\bibfnamefont{M.}~\bibnamefont{Lippmaa}},
  \bibinfo{journal}{ACS applied materials \& interfaces}
  \textbf{\bibinfo{volume}{12}}, \bibinfo{pages}{25042} (\bibinfo{year}{2020}).

\bibitem[{\citenamefont{Kong et~al.}(2018)\citenamefont{Kong, Li, Qiao, Kim,
  Lee, Nie, Lee, Osadchy, Molnar, Gaskill et~al.}}]{kong2018polarity}
\bibinfo{author}{\bibfnamefont{W.}~\bibnamefont{Kong}},
  \bibinfo{author}{\bibfnamefont{H.}~\bibnamefont{Li}},
  \bibinfo{author}{\bibfnamefont{K.}~\bibnamefont{Qiao}},
  \bibinfo{author}{\bibfnamefont{Y.}~\bibnamefont{Kim}},
  \bibinfo{author}{\bibfnamefont{K.}~\bibnamefont{Lee}},
  \bibinfo{author}{\bibfnamefont{Y.}~\bibnamefont{Nie}},
  \bibinfo{author}{\bibfnamefont{D.}~\bibnamefont{Lee}},
  \bibinfo{author}{\bibfnamefont{T.}~\bibnamefont{Osadchy}},
  \bibinfo{author}{\bibfnamefont{R.~J.} \bibnamefont{Molnar}},
  \bibinfo{author}{\bibfnamefont{D.~K.} \bibnamefont{Gaskill}},
  \bibnamefont{et~al.}, \bibinfo{journal}{Nature materials}
  \textbf{\bibinfo{volume}{17}}, \bibinfo{pages}{999} (\bibinfo{year}{2018}).

\bibitem[{\citenamefont{Manzo et~al.}(2022)\citenamefont{Manzo, Strohbeen, Lim,
  Saraswat, Du, Xu, Pokharel, Mawst, Arnold, and Kawasaki}}]{manzo2022pinhole}
\bibinfo{author}{\bibfnamefont{S.}~\bibnamefont{Manzo}},
  \bibinfo{author}{\bibfnamefont{P.~J.} \bibnamefont{Strohbeen}},
  \bibinfo{author}{\bibfnamefont{Z.~H.} \bibnamefont{Lim}},
  \bibinfo{author}{\bibfnamefont{V.}~\bibnamefont{Saraswat}},
  \bibinfo{author}{\bibfnamefont{D.}~\bibnamefont{Du}},
  \bibinfo{author}{\bibfnamefont{S.}~\bibnamefont{Xu}},
  \bibinfo{author}{\bibfnamefont{N.}~\bibnamefont{Pokharel}},
  \bibinfo{author}{\bibfnamefont{L.~J.} \bibnamefont{Mawst}},
  \bibinfo{author}{\bibfnamefont{M.~S.} \bibnamefont{Arnold}},
  \bibnamefont{and} \bibinfo{author}{\bibfnamefont{J.~K.}
  \bibnamefont{Kawasaki}}, \bibinfo{journal}{Nature communications}
  \textbf{\bibinfo{volume}{13}}, \bibinfo{pages}{1} (\bibinfo{year}{2022}).

\bibitem[{\citenamefont{Jiang et~al.}(2019)\citenamefont{Jiang, Sun, Chen,
  Wang, Chen, Hu, Guo, Zhang, Ma, Gao et~al.}}]{jiang2019carrier}
\bibinfo{author}{\bibfnamefont{J.}~\bibnamefont{Jiang}},
  \bibinfo{author}{\bibfnamefont{X.}~\bibnamefont{Sun}},
  \bibinfo{author}{\bibfnamefont{X.}~\bibnamefont{Chen}},
  \bibinfo{author}{\bibfnamefont{B.}~\bibnamefont{Wang}},
  \bibinfo{author}{\bibfnamefont{Z.}~\bibnamefont{Chen}},
  \bibinfo{author}{\bibfnamefont{Y.}~\bibnamefont{Hu}},
  \bibinfo{author}{\bibfnamefont{Y.}~\bibnamefont{Guo}},
  \bibinfo{author}{\bibfnamefont{L.}~\bibnamefont{Zhang}},
  \bibinfo{author}{\bibfnamefont{Y.}~\bibnamefont{Ma}},
  \bibinfo{author}{\bibfnamefont{L.}~\bibnamefont{Gao}}, \bibnamefont{et~al.},
  \bibinfo{journal}{Nature communications} \textbf{\bibinfo{volume}{10}},
  \bibinfo{pages}{1} (\bibinfo{year}{2019}).

\bibitem[{\citenamefont{Lu et~al.}(2018)\citenamefont{Lu, Sun, Xie, Littlejohn,
  Wang, Zhang, Washington, and Lu}}]{lu2018remote}
\bibinfo{author}{\bibfnamefont{Z.}~\bibnamefont{Lu}},
  \bibinfo{author}{\bibfnamefont{X.}~\bibnamefont{Sun}},
  \bibinfo{author}{\bibfnamefont{W.}~\bibnamefont{Xie}},
  \bibinfo{author}{\bibfnamefont{A.}~\bibnamefont{Littlejohn}},
  \bibinfo{author}{\bibfnamefont{G.-C.} \bibnamefont{Wang}},
  \bibinfo{author}{\bibfnamefont{S.}~\bibnamefont{Zhang}},
  \bibinfo{author}{\bibfnamefont{M.~A.} \bibnamefont{Washington}},
  \bibnamefont{and} \bibinfo{author}{\bibfnamefont{T.-M.} \bibnamefont{Lu}},
  \bibinfo{journal}{Nanotechnology} \textbf{\bibinfo{volume}{29}},
  \bibinfo{pages}{445702} (\bibinfo{year}{2018}).

\bibitem[{\citenamefont{Kim et~al.}(2021{\natexlab{a}})\citenamefont{Kim, Kim,
  Jeong, Yu, Lu, Lee, Kim, Jeong, Kim, and Kim}}]{kim2021role}
\bibinfo{author}{\bibfnamefont{H.}~\bibnamefont{Kim}},
  \bibinfo{author}{\bibfnamefont{J.~C.} \bibnamefont{Kim}},
  \bibinfo{author}{\bibfnamefont{Y.}~\bibnamefont{Jeong}},
  \bibinfo{author}{\bibfnamefont{J.}~\bibnamefont{Yu}},
  \bibinfo{author}{\bibfnamefont{K.}~\bibnamefont{Lu}},
  \bibinfo{author}{\bibfnamefont{D.}~\bibnamefont{Lee}},
  \bibinfo{author}{\bibfnamefont{N.}~\bibnamefont{Kim}},
  \bibinfo{author}{\bibfnamefont{H.~Y.} \bibnamefont{Jeong}},
  \bibinfo{author}{\bibfnamefont{J.}~\bibnamefont{Kim}}, \bibnamefont{and}
  \bibinfo{author}{\bibfnamefont{S.}~\bibnamefont{Kim}},
  \bibinfo{journal}{Journal of Applied Physics} \textbf{\bibinfo{volume}{130}},
  \bibinfo{pages}{174901} (\bibinfo{year}{2021}{\natexlab{a}}).

\bibitem[{\citenamefont{Kim et~al.}(2021{\natexlab{b}})\citenamefont{Kim, Lu,
  Liu, Kum, Kim, Qiao, Bae, Lee, Ji, Kim et~al.}}]{kim2021impact}
\bibinfo{author}{\bibfnamefont{H.}~\bibnamefont{Kim}},
  \bibinfo{author}{\bibfnamefont{K.}~\bibnamefont{Lu}},
  \bibinfo{author}{\bibfnamefont{Y.}~\bibnamefont{Liu}},
  \bibinfo{author}{\bibfnamefont{H.~S.} \bibnamefont{Kum}},
  \bibinfo{author}{\bibfnamefont{K.~S.} \bibnamefont{Kim}},
  \bibinfo{author}{\bibfnamefont{K.}~\bibnamefont{Qiao}},
  \bibinfo{author}{\bibfnamefont{S.-H.} \bibnamefont{Bae}},
  \bibinfo{author}{\bibfnamefont{S.}~\bibnamefont{Lee}},
  \bibinfo{author}{\bibfnamefont{Y.~J.} \bibnamefont{Ji}},
  \bibinfo{author}{\bibfnamefont{K.~H.} \bibnamefont{Kim}},
  \bibnamefont{et~al.}, \bibinfo{journal}{ACS nano}
  \textbf{\bibinfo{volume}{15}}, \bibinfo{pages}{10587}
  (\bibinfo{year}{2021}{\natexlab{b}}).

\bibitem[{\citenamefont{Harbola et~al.}(2021)\citenamefont{Harbola, Crossley,
  Hong, Lu, Birkholzer, Hikita, and Hwang}}]{harbola2021strain}
\bibinfo{author}{\bibfnamefont{V.}~\bibnamefont{Harbola}},
  \bibinfo{author}{\bibfnamefont{S.}~\bibnamefont{Crossley}},
  \bibinfo{author}{\bibfnamefont{S.~S.} \bibnamefont{Hong}},
  \bibinfo{author}{\bibfnamefont{D.}~\bibnamefont{Lu}},
  \bibinfo{author}{\bibfnamefont{Y.~A.} \bibnamefont{Birkholzer}},
  \bibinfo{author}{\bibfnamefont{Y.}~\bibnamefont{Hikita}}, \bibnamefont{and}
  \bibinfo{author}{\bibfnamefont{H.~Y.} \bibnamefont{Hwang}},
  \bibinfo{journal}{Nano letters} \textbf{\bibinfo{volume}{21}},
  \bibinfo{pages}{2470} (\bibinfo{year}{2021}).

\bibitem[{\citenamefont{Prakash et~al.}(2022)\citenamefont{Prakash, Chen,
  Debasu, Savage, Tangpatjaroen, Deneke, Malachias, Alfieri, Elleuch, Lekhal
  et~al.}}]{prakash2022reconfiguration}
\bibinfo{author}{\bibfnamefont{D.~J.} \bibnamefont{Prakash}},
  \bibinfo{author}{\bibfnamefont{Y.}~\bibnamefont{Chen}},
  \bibinfo{author}{\bibfnamefont{M.~L.} \bibnamefont{Debasu}},
  \bibinfo{author}{\bibfnamefont{D.~E.} \bibnamefont{Savage}},
  \bibinfo{author}{\bibfnamefont{C.}~\bibnamefont{Tangpatjaroen}},
  \bibinfo{author}{\bibfnamefont{C.}~\bibnamefont{Deneke}},
  \bibinfo{author}{\bibfnamefont{A.}~\bibnamefont{Malachias}},
  \bibinfo{author}{\bibfnamefont{A.~D.} \bibnamefont{Alfieri}},
  \bibinfo{author}{\bibfnamefont{O.}~\bibnamefont{Elleuch}},
  \bibinfo{author}{\bibfnamefont{K.}~\bibnamefont{Lekhal}},
  \bibnamefont{et~al.}, \bibinfo{journal}{Small} \textbf{\bibinfo{volume}{18}},
  \bibinfo{pages}{2105424} (\bibinfo{year}{2022}).

\bibitem[{\citenamefont{Xie et~al.}(2018)\citenamefont{Xie, Tu, Han, Huang,
  Kang, Lao, Poddar, Park, Muller, DiStasio~Jr et~al.}}]{xie2018coherent}
\bibinfo{author}{\bibfnamefont{S.}~\bibnamefont{Xie}},
  \bibinfo{author}{\bibfnamefont{L.}~\bibnamefont{Tu}},
  \bibinfo{author}{\bibfnamefont{Y.}~\bibnamefont{Han}},
  \bibinfo{author}{\bibfnamefont{L.}~\bibnamefont{Huang}},
  \bibinfo{author}{\bibfnamefont{K.}~\bibnamefont{Kang}},
  \bibinfo{author}{\bibfnamefont{K.~U.} \bibnamefont{Lao}},
  \bibinfo{author}{\bibfnamefont{P.}~\bibnamefont{Poddar}},
  \bibinfo{author}{\bibfnamefont{C.}~\bibnamefont{Park}},
  \bibinfo{author}{\bibfnamefont{D.~A.} \bibnamefont{Muller}},
  \bibinfo{author}{\bibfnamefont{R.~A.} \bibnamefont{DiStasio~Jr}},
  \bibnamefont{et~al.}, \bibinfo{journal}{Science}
  \textbf{\bibinfo{volume}{359}}, \bibinfo{pages}{1131} (\bibinfo{year}{2018}).

\bibitem[{\citenamefont{Lee}(2020)}]{lee2020flexoelectricity}
\bibinfo{author}{\bibfnamefont{D.}~\bibnamefont{Lee}}, \bibinfo{journal}{Apl
  Materials} \textbf{\bibinfo{volume}{8}}, \bibinfo{pages}{090901}
  (\bibinfo{year}{2020}).

\bibitem[{\citenamefont{Cai et~al.}(2022)\citenamefont{Cai, Lun, Ji, Lv, Han,
  Guo, Zang, Gao, Wei, Gu et~al.}}]{cai2022enhanced}
\bibinfo{author}{\bibfnamefont{S.}~\bibnamefont{Cai}},
  \bibinfo{author}{\bibfnamefont{Y.}~\bibnamefont{Lun}},
  \bibinfo{author}{\bibfnamefont{D.}~\bibnamefont{Ji}},
  \bibinfo{author}{\bibfnamefont{P.}~\bibnamefont{Lv}},
  \bibinfo{author}{\bibfnamefont{L.}~\bibnamefont{Han}},
  \bibinfo{author}{\bibfnamefont{C.}~\bibnamefont{Guo}},
  \bibinfo{author}{\bibfnamefont{Y.}~\bibnamefont{Zang}},
  \bibinfo{author}{\bibfnamefont{S.}~\bibnamefont{Gao}},
  \bibinfo{author}{\bibfnamefont{Y.}~\bibnamefont{Wei}},
  \bibinfo{author}{\bibfnamefont{M.}~\bibnamefont{Gu}}, \bibnamefont{et~al.}
  (\bibinfo{year}{2022}).

\bibitem[{\citenamefont{Schmidt and Eberl}(2001)}]{schmidt2001thin}
\bibinfo{author}{\bibfnamefont{O.~G.} \bibnamefont{Schmidt}} \bibnamefont{and}
  \bibinfo{author}{\bibfnamefont{K.}~\bibnamefont{Eberl}},
  \bibinfo{journal}{Nature} \textbf{\bibinfo{volume}{410}},
  \bibinfo{pages}{168} (\bibinfo{year}{2001}).

\bibitem[{\citenamefont{Huang et~al.}(2005)\citenamefont{Huang, Boone, Roberts,
  Savage, Lagally, Shaji, Qin, Blick, Nairn, and
  Liu}}]{huang2005nanomechanical}
\bibinfo{author}{\bibfnamefont{M.}~\bibnamefont{Huang}},
  \bibinfo{author}{\bibfnamefont{C.}~\bibnamefont{Boone}},
  \bibinfo{author}{\bibfnamefont{M.}~\bibnamefont{Roberts}},
  \bibinfo{author}{\bibfnamefont{D.~E.} \bibnamefont{Savage}},
  \bibinfo{author}{\bibfnamefont{M.~G.} \bibnamefont{Lagally}},
  \bibinfo{author}{\bibfnamefont{N.}~\bibnamefont{Shaji}},
  \bibinfo{author}{\bibfnamefont{H.}~\bibnamefont{Qin}},
  \bibinfo{author}{\bibfnamefont{R.}~\bibnamefont{Blick}},
  \bibinfo{author}{\bibfnamefont{J.~A.} \bibnamefont{Nairn}}, \bibnamefont{and}
  \bibinfo{author}{\bibfnamefont{F.}~\bibnamefont{Liu}},
  \bibinfo{journal}{Advanced Materials} \textbf{\bibinfo{volume}{17}},
  \bibinfo{pages}{2860} (\bibinfo{year}{2005}).

\bibitem[{\citenamefont{Zangwill}(1988)}]{zangwill1988physics}
\bibinfo{author}{\bibfnamefont{A.}~\bibnamefont{Zangwill}},
  \emph{\bibinfo{title}{Physics at surfaces}} (\bibinfo{publisher}{Cambridge
  university press}, \bibinfo{year}{1988}).

\bibitem[{\citenamefont{Sears et~al.}(1954)\citenamefont{Sears, Gatti, and
  Fullman}}]{sears1954elastic}
\bibinfo{author}{\bibfnamefont{G.}~\bibnamefont{Sears}},
  \bibinfo{author}{\bibfnamefont{A.}~\bibnamefont{Gatti}}, \bibnamefont{and}
  \bibinfo{author}{\bibfnamefont{R.}~\bibnamefont{Fullman}},
  \bibinfo{journal}{Acta Metallurgica} \textbf{\bibinfo{volume}{2}},
  \bibinfo{pages}{727} (\bibinfo{year}{1954}).

\bibitem[{\citenamefont{Herring and Galt}(1952)}]{herring1952elastic}
\bibinfo{author}{\bibfnamefont{C.}~\bibnamefont{Herring}} \bibnamefont{and}
  \bibinfo{author}{\bibfnamefont{J.}~\bibnamefont{Galt}},
  \bibinfo{journal}{Physical Review} \textbf{\bibinfo{volume}{85}},
  \bibinfo{pages}{1060} (\bibinfo{year}{1952}).

\bibitem[{\citenamefont{Brenner}(1956)}]{brenner1956tensile}
\bibinfo{author}{\bibfnamefont{S.~S.} \bibnamefont{Brenner}},
  \bibinfo{journal}{Journal of applied physics} \textbf{\bibinfo{volume}{27}},
  \bibinfo{pages}{1484} (\bibinfo{year}{1956}).

\bibitem[{\citenamefont{Wang et~al.}(2011)\citenamefont{Wang, Wang, Joyce, Gao,
  Liao, Mai, Tan, Zou, Ringer, Gao et~al.}}]{wang2011super}
\bibinfo{author}{\bibfnamefont{Y.-B.} \bibnamefont{Wang}},
  \bibinfo{author}{\bibfnamefont{L.-F.} \bibnamefont{Wang}},
  \bibinfo{author}{\bibfnamefont{H.~J.} \bibnamefont{Joyce}},
  \bibinfo{author}{\bibfnamefont{Q.}~\bibnamefont{Gao}},
  \bibinfo{author}{\bibfnamefont{X.-Z.} \bibnamefont{Liao}},
  \bibinfo{author}{\bibfnamefont{Y.-W.} \bibnamefont{Mai}},
  \bibinfo{author}{\bibfnamefont{H.~H.} \bibnamefont{Tan}},
  \bibinfo{author}{\bibfnamefont{J.}~\bibnamefont{Zou}},
  \bibinfo{author}{\bibfnamefont{S.~P.} \bibnamefont{Ringer}},
  \bibinfo{author}{\bibfnamefont{H.-J.} \bibnamefont{Gao}},
  \bibnamefont{et~al.}, \bibinfo{journal}{Advanced Materials}
  \textbf{\bibinfo{volume}{23}}, \bibinfo{pages}{1356} (\bibinfo{year}{2011}).

\bibitem[{\citenamefont{Wang et~al.}(2017)\citenamefont{Wang, Shan, and
  Huang}}]{wang2017mechanical}
\bibinfo{author}{\bibfnamefont{S.}~\bibnamefont{Wang}},
  \bibinfo{author}{\bibfnamefont{Z.}~\bibnamefont{Shan}}, \bibnamefont{and}
  \bibinfo{author}{\bibfnamefont{H.}~\bibnamefont{Huang}},
  \bibinfo{journal}{Advanced Science} \textbf{\bibinfo{volume}{4}},
  \bibinfo{pages}{1600332} (\bibinfo{year}{2017}).

\bibitem[{\citenamefont{Yue et~al.}(2011)\citenamefont{Yue, Liu, Zhang, Han,
  and Ma}}]{yue2011approaching}
\bibinfo{author}{\bibfnamefont{Y.}~\bibnamefont{Yue}},
  \bibinfo{author}{\bibfnamefont{P.}~\bibnamefont{Liu}},
  \bibinfo{author}{\bibfnamefont{Z.}~\bibnamefont{Zhang}},
  \bibinfo{author}{\bibfnamefont{X.}~\bibnamefont{Han}}, \bibnamefont{and}
  \bibinfo{author}{\bibfnamefont{E.}~\bibnamefont{Ma}}, \bibinfo{journal}{Nano
  letters} \textbf{\bibinfo{volume}{11}}, \bibinfo{pages}{3151}
  (\bibinfo{year}{2011}).

\bibitem[{\citenamefont{Li et~al.}(2021)\citenamefont{Li, Chu, Liu, Jiang, Qu,
  Gao, Wang, Ren, Sun, Chen et~al.}}]{li2021superelastic}
\bibinfo{author}{\bibfnamefont{Y.}~\bibnamefont{Li}},
  \bibinfo{author}{\bibfnamefont{K.}~\bibnamefont{Chu}},
  \bibinfo{author}{\bibfnamefont{C.}~\bibnamefont{Liu}},
  \bibinfo{author}{\bibfnamefont{P.}~\bibnamefont{Jiang}},
  \bibinfo{author}{\bibfnamefont{K.}~\bibnamefont{Qu}},
  \bibinfo{author}{\bibfnamefont{P.}~\bibnamefont{Gao}},
  \bibinfo{author}{\bibfnamefont{J.}~\bibnamefont{Wang}},
  \bibinfo{author}{\bibfnamefont{F.}~\bibnamefont{Ren}},
  \bibinfo{author}{\bibfnamefont{Q.}~\bibnamefont{Sun}},
  \bibinfo{author}{\bibfnamefont{L.}~\bibnamefont{Chen}}, \bibnamefont{et~al.},
  \bibinfo{journal}{Proceedings of the National Academy of Sciences}
  \textbf{\bibinfo{volume}{118}}, \bibinfo{pages}{e2025255118}
  (\bibinfo{year}{2021}).

\bibitem[{\citenamefont{Deng et~al.}(2019)\citenamefont{Deng, Gammer, Ciston,
  Ercius, Ophus, Bustillo, Song, Zhang, Wu, Du
  et~al.}}]{deng2019hierarchically}
\bibinfo{author}{\bibfnamefont{Y.}~\bibnamefont{Deng}},
  \bibinfo{author}{\bibfnamefont{C.}~\bibnamefont{Gammer}},
  \bibinfo{author}{\bibfnamefont{J.}~\bibnamefont{Ciston}},
  \bibinfo{author}{\bibfnamefont{P.}~\bibnamefont{Ercius}},
  \bibinfo{author}{\bibfnamefont{C.}~\bibnamefont{Ophus}},
  \bibinfo{author}{\bibfnamefont{K.}~\bibnamefont{Bustillo}},
  \bibinfo{author}{\bibfnamefont{C.}~\bibnamefont{Song}},
  \bibinfo{author}{\bibfnamefont{R.}~\bibnamefont{Zhang}},
  \bibinfo{author}{\bibfnamefont{D.}~\bibnamefont{Wu}},
  \bibinfo{author}{\bibfnamefont{Y.}~\bibnamefont{Du}}, \bibnamefont{et~al.},
  \bibinfo{journal}{Acta Materialia} \textbf{\bibinfo{volume}{181}},
  \bibinfo{pages}{501} (\bibinfo{year}{2019}).

\bibitem[{\citenamefont{M{\"u}ller and Sa{\'u}l}(2004)}]{muller2004elastic}
\bibinfo{author}{\bibfnamefont{P.}~\bibnamefont{M{\"u}ller}} \bibnamefont{and}
  \bibinfo{author}{\bibfnamefont{A.}~\bibnamefont{Sa{\'u}l}},
  \bibinfo{journal}{Surface Science Reports} \textbf{\bibinfo{volume}{54}},
  \bibinfo{pages}{157} (\bibinfo{year}{2004}).

\bibitem[{\citenamefont{Liu et~al.}(2009)\citenamefont{Liu, Li, Zhou, Yang, Ma,
  Xie, Pan, and Sun}}]{liu2009size}
\bibinfo{author}{\bibfnamefont{X.}~\bibnamefont{Liu}},
  \bibinfo{author}{\bibfnamefont{J.}~\bibnamefont{Li}},
  \bibinfo{author}{\bibfnamefont{Z.}~\bibnamefont{Zhou}},
  \bibinfo{author}{\bibfnamefont{L.}~\bibnamefont{Yang}},
  \bibinfo{author}{\bibfnamefont{Z.}~\bibnamefont{Ma}},
  \bibinfo{author}{\bibfnamefont{G.}~\bibnamefont{Xie}},
  \bibinfo{author}{\bibfnamefont{Y.}~\bibnamefont{Pan}}, \bibnamefont{and}
  \bibinfo{author}{\bibfnamefont{C.~Q.} \bibnamefont{Sun}},
  \bibinfo{journal}{Applied Physics Letters} \textbf{\bibinfo{volume}{94}},
  \bibinfo{pages}{131902} (\bibinfo{year}{2009}).

\bibitem[{\citenamefont{Wang et~al.}(2013)\citenamefont{Wang, Liu, Guan, Yang,
  Sun, Cheng, Hirata, Zhang, Ma, Chen et~al.}}]{wang2013situ}
\bibinfo{author}{\bibfnamefont{L.}~\bibnamefont{Wang}},
  \bibinfo{author}{\bibfnamefont{P.}~\bibnamefont{Liu}},
  \bibinfo{author}{\bibfnamefont{P.}~\bibnamefont{Guan}},
  \bibinfo{author}{\bibfnamefont{M.}~\bibnamefont{Yang}},
  \bibinfo{author}{\bibfnamefont{J.}~\bibnamefont{Sun}},
  \bibinfo{author}{\bibfnamefont{Y.}~\bibnamefont{Cheng}},
  \bibinfo{author}{\bibfnamefont{A.}~\bibnamefont{Hirata}},
  \bibinfo{author}{\bibfnamefont{Z.}~\bibnamefont{Zhang}},
  \bibinfo{author}{\bibfnamefont{E.}~\bibnamefont{Ma}},
  \bibinfo{author}{\bibfnamefont{M.}~\bibnamefont{Chen}}, \bibnamefont{et~al.},
  \bibinfo{journal}{Nature communications} \textbf{\bibinfo{volume}{4}},
  \bibinfo{pages}{1} (\bibinfo{year}{2013}).

\bibitem[{\citenamefont{Uchic et~al.}(2004)\citenamefont{Uchic, Dimiduk,
  Florando, and Nix}}]{uchic2004sample}
\bibinfo{author}{\bibfnamefont{M.~D.} \bibnamefont{Uchic}},
  \bibinfo{author}{\bibfnamefont{D.~M.} \bibnamefont{Dimiduk}},
  \bibinfo{author}{\bibfnamefont{J.~N.} \bibnamefont{Florando}},
  \bibnamefont{and} \bibinfo{author}{\bibfnamefont{W.~D.} \bibnamefont{Nix}},
  \bibinfo{journal}{Science} \textbf{\bibinfo{volume}{305}},
  \bibinfo{pages}{986} (\bibinfo{year}{2004}).

\bibitem[{\citenamefont{Stengel and Vanderbilt}(2017)}]{stengel2017first}
\bibinfo{author}{\bibfnamefont{M.}~\bibnamefont{Stengel}} \bibnamefont{and}
  \bibinfo{author}{\bibfnamefont{D.}~\bibnamefont{Vanderbilt}}, in
  \emph{\bibinfo{booktitle}{Flexoelectricity In Solids: From Theory To
  Applications}} (\bibinfo{publisher}{World Scientific}, \bibinfo{year}{2017}),
  pp. \bibinfo{pages}{31--110}.

\bibitem[{\citenamefont{Dreyer et~al.}(2018)\citenamefont{Dreyer, Stengel, and
  Vanderbilt}}]{dreyer2018current}
\bibinfo{author}{\bibfnamefont{C.~E.} \bibnamefont{Dreyer}},
  \bibinfo{author}{\bibfnamefont{M.}~\bibnamefont{Stengel}}, \bibnamefont{and}
  \bibinfo{author}{\bibfnamefont{D.}~\bibnamefont{Vanderbilt}},
  \bibinfo{journal}{Physical Review B} \textbf{\bibinfo{volume}{98}},
  \bibinfo{pages}{075153} (\bibinfo{year}{2018}).

\bibitem[{\citenamefont{Schiaffino et~al.}(2019)\citenamefont{Schiaffino,
  Dreyer, Vanderbilt, and Stengel}}]{schiaffino2019metric}
\bibinfo{author}{\bibfnamefont{A.}~\bibnamefont{Schiaffino}},
  \bibinfo{author}{\bibfnamefont{C.~E.} \bibnamefont{Dreyer}},
  \bibinfo{author}{\bibfnamefont{D.}~\bibnamefont{Vanderbilt}},
  \bibnamefont{and} \bibinfo{author}{\bibfnamefont{M.}~\bibnamefont{Stengel}},
  \bibinfo{journal}{Physical Review B} \textbf{\bibinfo{volume}{99}},
  \bibinfo{pages}{085107} (\bibinfo{year}{2019}).

\bibitem[{\citenamefont{Royo and Stengel}(2019)}]{royo2019first}
\bibinfo{author}{\bibfnamefont{M.}~\bibnamefont{Royo}} \bibnamefont{and}
  \bibinfo{author}{\bibfnamefont{M.}~\bibnamefont{Stengel}},
  \bibinfo{journal}{Physical Review X} \textbf{\bibinfo{volume}{9}},
  \bibinfo{pages}{021050} (\bibinfo{year}{2019}).

\bibitem[{\citenamefont{Springolo et~al.}(2021)\citenamefont{Springolo, Royo,
  and Stengel}}]{springolo2021direct}
\bibinfo{author}{\bibfnamefont{M.}~\bibnamefont{Springolo}},
  \bibinfo{author}{\bibfnamefont{M.}~\bibnamefont{Royo}}, \bibnamefont{and}
  \bibinfo{author}{\bibfnamefont{M.}~\bibnamefont{Stengel}},
  \bibinfo{journal}{Physical review letters} \textbf{\bibinfo{volume}{127}},
  \bibinfo{pages}{216801} (\bibinfo{year}{2021}).

\bibitem[{\citenamefont{Royo and Stengel}(2022)}]{royo2022lattice}
\bibinfo{author}{\bibfnamefont{M.}~\bibnamefont{Royo}} \bibnamefont{and}
  \bibinfo{author}{\bibfnamefont{M.}~\bibnamefont{Stengel}},
  \bibinfo{journal}{Physical Review B} \textbf{\bibinfo{volume}{105}},
  \bibinfo{pages}{064101} (\bibinfo{year}{2022}).

\bibitem[{\citenamefont{Eliseev et~al.}(2012)\citenamefont{Eliseev, Morozovska,
  Gu, Borisevich, Chen, Gopalan, and Kalinin}}]{eliseev2012conductivity}
\bibinfo{author}{\bibfnamefont{E.~A.} \bibnamefont{Eliseev}},
  \bibinfo{author}{\bibfnamefont{A.~N.} \bibnamefont{Morozovska}},
  \bibinfo{author}{\bibfnamefont{Y.}~\bibnamefont{Gu}},
  \bibinfo{author}{\bibfnamefont{A.~Y.} \bibnamefont{Borisevich}},
  \bibinfo{author}{\bibfnamefont{L.-Q.} \bibnamefont{Chen}},
  \bibinfo{author}{\bibfnamefont{V.}~\bibnamefont{Gopalan}}, \bibnamefont{and}
  \bibinfo{author}{\bibfnamefont{S.~V.} \bibnamefont{Kalinin}},
  \bibinfo{journal}{Physical Review B} \textbf{\bibinfo{volume}{86}},
  \bibinfo{pages}{085416} (\bibinfo{year}{2012}).

\bibitem[{\citenamefont{Stolichnov et~al.}(2015)\citenamefont{Stolichnov,
  Feigl, McGilly, Sluka, Wei, Colla, Crassous, Shapovalov, Yudin, Tagantsev
  et~al.}}]{stolichnov2015bent}
\bibinfo{author}{\bibfnamefont{I.}~\bibnamefont{Stolichnov}},
  \bibinfo{author}{\bibfnamefont{L.}~\bibnamefont{Feigl}},
  \bibinfo{author}{\bibfnamefont{L.~J.} \bibnamefont{McGilly}},
  \bibinfo{author}{\bibfnamefont{T.}~\bibnamefont{Sluka}},
  \bibinfo{author}{\bibfnamefont{X.-K.} \bibnamefont{Wei}},
  \bibinfo{author}{\bibfnamefont{E.}~\bibnamefont{Colla}},
  \bibinfo{author}{\bibfnamefont{A.}~\bibnamefont{Crassous}},
  \bibinfo{author}{\bibfnamefont{K.}~\bibnamefont{Shapovalov}},
  \bibinfo{author}{\bibfnamefont{P.}~\bibnamefont{Yudin}},
  \bibinfo{author}{\bibfnamefont{A.~K.} \bibnamefont{Tagantsev}},
  \bibnamefont{et~al.}, \bibinfo{journal}{Nano Letters}
  \textbf{\bibinfo{volume}{15}}, \bibinfo{pages}{8049} (\bibinfo{year}{2015}).

\bibitem[{\citenamefont{Zhong et~al.}(1994)\citenamefont{Zhong, Vanderbilt, and
  Rabe}}]{zhong1994phase}
\bibinfo{author}{\bibfnamefont{W.}~\bibnamefont{Zhong}},
  \bibinfo{author}{\bibfnamefont{D.}~\bibnamefont{Vanderbilt}},
  \bibnamefont{and} \bibinfo{author}{\bibfnamefont{K.}~\bibnamefont{Rabe}},
  \bibinfo{journal}{Physical Review Letters} \textbf{\bibinfo{volume}{73}},
  \bibinfo{pages}{1861} (\bibinfo{year}{1994}).

\bibitem[{\citenamefont{Bellaiche et~al.}(2000)\citenamefont{Bellaiche,
  Garc{\'\i}a, and Vanderbilt}}]{bellaiche2000finite}
\bibinfo{author}{\bibfnamefont{L.}~\bibnamefont{Bellaiche}},
  \bibinfo{author}{\bibfnamefont{A.}~\bibnamefont{Garc{\'\i}a}},
  \bibnamefont{and}
  \bibinfo{author}{\bibfnamefont{D.}~\bibnamefont{Vanderbilt}},
  \bibinfo{journal}{Physical Review Letters} \textbf{\bibinfo{volume}{84}},
  \bibinfo{pages}{5427} (\bibinfo{year}{2000}).

\bibitem[{\citenamefont{Lai et~al.}(2006)\citenamefont{Lai, Ponomareva, Naumov,
  Kornev, Fu, Bellaiche, and Salamo}}]{lai2006electric}
\bibinfo{author}{\bibfnamefont{B.-K.} \bibnamefont{Lai}},
  \bibinfo{author}{\bibfnamefont{I.}~\bibnamefont{Ponomareva}},
  \bibinfo{author}{\bibfnamefont{I.}~\bibnamefont{Naumov}},
  \bibinfo{author}{\bibfnamefont{I.}~\bibnamefont{Kornev}},
  \bibinfo{author}{\bibfnamefont{H.}~\bibnamefont{Fu}},
  \bibinfo{author}{\bibfnamefont{L.}~\bibnamefont{Bellaiche}},
  \bibnamefont{and} \bibinfo{author}{\bibfnamefont{G.}~\bibnamefont{Salamo}},
  \bibinfo{journal}{Physical review letters} \textbf{\bibinfo{volume}{96}},
  \bibinfo{pages}{137602} (\bibinfo{year}{2006}).

\bibitem[{\citenamefont{Nahas et~al.}(2020)\citenamefont{Nahas, Prokhorenko,
  Fischer, Xu, Carr{\'e}t{\'e}ro, Prosandeev, Bibes, Fusil, Dkhil, Garcia
  et~al.}}]{nahas2020inverse}
\bibinfo{author}{\bibfnamefont{Y.}~\bibnamefont{Nahas}},
  \bibinfo{author}{\bibfnamefont{S.}~\bibnamefont{Prokhorenko}},
  \bibinfo{author}{\bibfnamefont{J.}~\bibnamefont{Fischer}},
  \bibinfo{author}{\bibfnamefont{B.}~\bibnamefont{Xu}},
  \bibinfo{author}{\bibfnamefont{C.}~\bibnamefont{Carr{\'e}t{\'e}ro}},
  \bibinfo{author}{\bibfnamefont{S.}~\bibnamefont{Prosandeev}},
  \bibinfo{author}{\bibfnamefont{M.}~\bibnamefont{Bibes}},
  \bibinfo{author}{\bibfnamefont{S.}~\bibnamefont{Fusil}},
  \bibinfo{author}{\bibfnamefont{B.}~\bibnamefont{Dkhil}},
  \bibinfo{author}{\bibfnamefont{V.}~\bibnamefont{Garcia}},
  \bibnamefont{et~al.}, \bibinfo{journal}{Nature}
  \textbf{\bibinfo{volume}{577}}, \bibinfo{pages}{47} (\bibinfo{year}{2020}).

\bibitem[{\citenamefont{Wojde{\l} et~al.}(2013)\citenamefont{Wojde{\l}, Hermet,
  Ljungberg, Ghosez, and Iniguez}}]{wojdel2013first}
\bibinfo{author}{\bibfnamefont{J.~C.} \bibnamefont{Wojde{\l}}},
  \bibinfo{author}{\bibfnamefont{P.}~\bibnamefont{Hermet}},
  \bibinfo{author}{\bibfnamefont{M.~P.} \bibnamefont{Ljungberg}},
  \bibinfo{author}{\bibfnamefont{P.}~\bibnamefont{Ghosez}}, \bibnamefont{and}
  \bibinfo{author}{\bibfnamefont{J.}~\bibnamefont{Iniguez}},
  \bibinfo{journal}{Journal of Physics: Condensed Matter}
  \textbf{\bibinfo{volume}{25}}, \bibinfo{pages}{305401}
  (\bibinfo{year}{2013}).

\bibitem[{\citenamefont{Wojde{\l} and Iniguez}(2014)}]{wojdel2014ferroelectric}
\bibinfo{author}{\bibfnamefont{J.~C.} \bibnamefont{Wojde{\l}}}
  \bibnamefont{and} \bibinfo{author}{\bibfnamefont{J.}~\bibnamefont{Iniguez}},
  \bibinfo{journal}{Physical Review Letters} \textbf{\bibinfo{volume}{112}},
  \bibinfo{pages}{247603} (\bibinfo{year}{2014}).

\bibitem[{\citenamefont{Garcia-Fernandez
  et~al.}(2016)\citenamefont{Garcia-Fernandez, Wojde{\l}, Iniguez, and
  Junquera}}]{garcia2016second}
\bibinfo{author}{\bibfnamefont{P.}~\bibnamefont{Garcia-Fernandez}},
  \bibinfo{author}{\bibfnamefont{J.~C.} \bibnamefont{Wojde{\l}}},
  \bibinfo{author}{\bibfnamefont{J.}~\bibnamefont{Iniguez}}, \bibnamefont{and}
  \bibinfo{author}{\bibfnamefont{J.}~\bibnamefont{Junquera}},
  \bibinfo{journal}{Physical Review B} \textbf{\bibinfo{volume}{93}},
  \bibinfo{pages}{195137} (\bibinfo{year}{2016}).

\bibitem[{\citenamefont{Park et~al.}(2022)\citenamefont{Park, Wang, Yang, Kim,
  Saremi, Zhao, Guzelturk, Sood, Nyby, Zajac et~al.}}]{park2022light}
\bibinfo{author}{\bibfnamefont{S.}~\bibnamefont{Park}},
  \bibinfo{author}{\bibfnamefont{B.}~\bibnamefont{Wang}},
  \bibinfo{author}{\bibfnamefont{T.}~\bibnamefont{Yang}},
  \bibinfo{author}{\bibfnamefont{J.}~\bibnamefont{Kim}},
  \bibinfo{author}{\bibfnamefont{S.}~\bibnamefont{Saremi}},
  \bibinfo{author}{\bibfnamefont{W.}~\bibnamefont{Zhao}},
  \bibinfo{author}{\bibfnamefont{B.}~\bibnamefont{Guzelturk}},
  \bibinfo{author}{\bibfnamefont{A.}~\bibnamefont{Sood}},
  \bibinfo{author}{\bibfnamefont{C.}~\bibnamefont{Nyby}},
  \bibinfo{author}{\bibfnamefont{M.}~\bibnamefont{Zajac}},
  \bibnamefont{et~al.}, \bibinfo{journal}{Nano Letters}
  (\bibinfo{year}{2022}).

\bibitem[{\citenamefont{Zhuo et~al.}(2022)\citenamefont{Zhuo, Zhou, Gao,
  H{\"o}fling, Dietrich, Groszewicz, Fulanovi{\'c}, Breckner, Wohninsland, Xu
  et~al.}}]{zhuo2022anisotropic}
\bibinfo{author}{\bibfnamefont{F.}~\bibnamefont{Zhuo}},
  \bibinfo{author}{\bibfnamefont{X.}~\bibnamefont{Zhou}},
  \bibinfo{author}{\bibfnamefont{S.}~\bibnamefont{Gao}},
  \bibinfo{author}{\bibfnamefont{M.}~\bibnamefont{H{\"o}fling}},
  \bibinfo{author}{\bibfnamefont{F.}~\bibnamefont{Dietrich}},
  \bibinfo{author}{\bibfnamefont{P.~B.} \bibnamefont{Groszewicz}},
  \bibinfo{author}{\bibfnamefont{L.}~\bibnamefont{Fulanovi{\'c}}},
  \bibinfo{author}{\bibfnamefont{P.}~\bibnamefont{Breckner}},
  \bibinfo{author}{\bibfnamefont{A.}~\bibnamefont{Wohninsland}},
  \bibinfo{author}{\bibfnamefont{B.-X.} \bibnamefont{Xu}},
  \bibnamefont{et~al.}, \bibinfo{journal}{Nature communications}
  \textbf{\bibinfo{volume}{13}}, \bibinfo{pages}{1} (\bibinfo{year}{2022}).

\bibitem[{\citenamefont{Dai et~al.}(2021)\citenamefont{Dai, Demirel, Liang, and
  Hu}}]{dai2021graph}
\bibinfo{author}{\bibfnamefont{M.}~\bibnamefont{Dai}},
  \bibinfo{author}{\bibfnamefont{M.~F.} \bibnamefont{Demirel}},
  \bibinfo{author}{\bibfnamefont{Y.}~\bibnamefont{Liang}}, \bibnamefont{and}
  \bibinfo{author}{\bibfnamefont{J.-M.} \bibnamefont{Hu}},
  \bibinfo{journal}{npj Computational Materials} \textbf{\bibinfo{volume}{7}},
  \bibinfo{pages}{1} (\bibinfo{year}{2021}).

\bibitem[{\citenamefont{Choudhury et~al.}(2007)\citenamefont{Choudhury, Li,
  Krill~Iii, and Chen}}]{choudhury2007effect}
\bibinfo{author}{\bibfnamefont{S.}~\bibnamefont{Choudhury}},
  \bibinfo{author}{\bibfnamefont{Y.}~\bibnamefont{Li}},
  \bibinfo{author}{\bibfnamefont{C.}~\bibnamefont{Krill~Iii}},
  \bibnamefont{and} \bibinfo{author}{\bibfnamefont{L.}~\bibnamefont{Chen}},
  \bibinfo{journal}{Acta materialia} \textbf{\bibinfo{volume}{55}},
  \bibinfo{pages}{1415} (\bibinfo{year}{2007}).

\bibitem[{\citenamefont{Hu et~al.}(2019)\citenamefont{Hu, Zhang, Yang, Shi,
  Cheng, Ni, Hao, Rao, and Chen}}]{hu2019phase}
\bibinfo{author}{\bibfnamefont{C.-C.} \bibnamefont{Hu}},
  \bibinfo{author}{\bibfnamefont{Z.}~\bibnamefont{Zhang}},
  \bibinfo{author}{\bibfnamefont{T.-N.} \bibnamefont{Yang}},
  \bibinfo{author}{\bibfnamefont{Y.-G.} \bibnamefont{Shi}},
  \bibinfo{author}{\bibfnamefont{X.-X.} \bibnamefont{Cheng}},
  \bibinfo{author}{\bibfnamefont{J.-J.} \bibnamefont{Ni}},
  \bibinfo{author}{\bibfnamefont{J.-G.} \bibnamefont{Hao}},
  \bibinfo{author}{\bibfnamefont{W.-F.} \bibnamefont{Rao}}, \bibnamefont{and}
  \bibinfo{author}{\bibfnamefont{L.-Q.} \bibnamefont{Chen}},
  \bibinfo{journal}{Applied physics letters} \textbf{\bibinfo{volume}{115}},
  \bibinfo{pages}{162402} (\bibinfo{year}{2019}).

\bibitem[{\citenamefont{Zhao et~al.}(2021)\citenamefont{Zhao, Gao, Yang,
  Scherer, Schulthei{\ss}, Meier, Tan, Kleebe, Chen, Koruza
  et~al.}}]{zhao2021precipitation}
\bibinfo{author}{\bibfnamefont{C.}~\bibnamefont{Zhao}},
  \bibinfo{author}{\bibfnamefont{S.}~\bibnamefont{Gao}},
  \bibinfo{author}{\bibfnamefont{T.}~\bibnamefont{Yang}},
  \bibinfo{author}{\bibfnamefont{M.}~\bibnamefont{Scherer}},
  \bibinfo{author}{\bibfnamefont{J.}~\bibnamefont{Schulthei{\ss}}},
  \bibinfo{author}{\bibfnamefont{D.}~\bibnamefont{Meier}},
  \bibinfo{author}{\bibfnamefont{X.}~\bibnamefont{Tan}},
  \bibinfo{author}{\bibfnamefont{H.-J.} \bibnamefont{Kleebe}},
  \bibinfo{author}{\bibfnamefont{L.-Q.} \bibnamefont{Chen}},
  \bibinfo{author}{\bibfnamefont{J.}~\bibnamefont{Koruza}},
  \bibnamefont{et~al.}, \bibinfo{journal}{Advanced Materials}
  \textbf{\bibinfo{volume}{33}}, \bibinfo{pages}{2102421}
  (\bibinfo{year}{2021}).

\bibitem[{\citenamefont{Liu et~al.}(2018)\citenamefont{Liu, Liu, Biegalski, Hu,
  Shang, Ji, Wang, Hsu, Wong, Cordill et~al.}}]{liu2018electrically}
\bibinfo{author}{\bibfnamefont{Z.}~\bibnamefont{Liu}},
  \bibinfo{author}{\bibfnamefont{J.}~\bibnamefont{Liu}},
  \bibinfo{author}{\bibfnamefont{M.~D.} \bibnamefont{Biegalski}},
  \bibinfo{author}{\bibfnamefont{J.-M.} \bibnamefont{Hu}},
  \bibinfo{author}{\bibfnamefont{S.}~\bibnamefont{Shang}},
  \bibinfo{author}{\bibfnamefont{Y.}~\bibnamefont{Ji}},
  \bibinfo{author}{\bibfnamefont{J.}~\bibnamefont{Wang}},
  \bibinfo{author}{\bibfnamefont{S.}~\bibnamefont{Hsu}},
  \bibinfo{author}{\bibfnamefont{A.}~\bibnamefont{Wong}},
  \bibinfo{author}{\bibfnamefont{M.}~\bibnamefont{Cordill}},
  \bibnamefont{et~al.}, \bibinfo{journal}{Nature communications}
  \textbf{\bibinfo{volume}{9}}, \bibinfo{pages}{1} (\bibinfo{year}{2018}).

\bibitem[{\citenamefont{Ohmer et~al.}(2022)\citenamefont{Ohmer, Yi, Gutfleisch,
  and Xu}}]{ohmer2022phase}
\bibinfo{author}{\bibfnamefont{D.}~\bibnamefont{Ohmer}},
  \bibinfo{author}{\bibfnamefont{M.}~\bibnamefont{Yi}},
  \bibinfo{author}{\bibfnamefont{O.}~\bibnamefont{Gutfleisch}},
  \bibnamefont{and} \bibinfo{author}{\bibfnamefont{B.-X.} \bibnamefont{Xu}},
  \bibinfo{journal}{International Journal of Solids and Structures}
  \textbf{\bibinfo{volume}{238}}, \bibinfo{pages}{111365}
  (\bibinfo{year}{2022}).

\bibitem[{\citenamefont{Huang et~al.}(2015)\citenamefont{Huang, Ma, Wang, Liu,
  He, and Chen}}]{huang2015phase}
\bibinfo{author}{\bibfnamefont{H.}~\bibnamefont{Huang}},
  \bibinfo{author}{\bibfnamefont{X.}~\bibnamefont{Ma}},
  \bibinfo{author}{\bibfnamefont{J.}~\bibnamefont{Wang}},
  \bibinfo{author}{\bibfnamefont{Z.}~\bibnamefont{Liu}},
  \bibinfo{author}{\bibfnamefont{W.}~\bibnamefont{He}}, \bibnamefont{and}
  \bibinfo{author}{\bibfnamefont{L.}~\bibnamefont{Chen}},
  \bibinfo{journal}{Acta Materialia} \textbf{\bibinfo{volume}{83}},
  \bibinfo{pages}{333} (\bibinfo{year}{2015}).

\bibitem[{\citenamefont{Wu et~al.}(2011)\citenamefont{Wu, Ma, Zhang, and
  Chen}}]{wu2011phase}
\bibinfo{author}{\bibfnamefont{P.}~\bibnamefont{Wu}},
  \bibinfo{author}{\bibfnamefont{X.}~\bibnamefont{Ma}},
  \bibinfo{author}{\bibfnamefont{J.}~\bibnamefont{Zhang}}, \bibnamefont{and}
  \bibinfo{author}{\bibfnamefont{L.}~\bibnamefont{Chen}},
  \bibinfo{journal}{Philosophical Magazine} \textbf{\bibinfo{volume}{91}},
  \bibinfo{pages}{2102} (\bibinfo{year}{2011}).

\bibitem[{\citenamefont{Qiu et~al.}(2020)\citenamefont{Qiu, Wang, Zhang, Zhang,
  Liu, Walker, Wang, Tian, Shrout, Xu et~al.}}]{qiu2020transparent}
\bibinfo{author}{\bibfnamefont{C.}~\bibnamefont{Qiu}},
  \bibinfo{author}{\bibfnamefont{B.}~\bibnamefont{Wang}},
  \bibinfo{author}{\bibfnamefont{N.}~\bibnamefont{Zhang}},
  \bibinfo{author}{\bibfnamefont{S.}~\bibnamefont{Zhang}},
  \bibinfo{author}{\bibfnamefont{J.}~\bibnamefont{Liu}},
  \bibinfo{author}{\bibfnamefont{D.}~\bibnamefont{Walker}},
  \bibinfo{author}{\bibfnamefont{Y.}~\bibnamefont{Wang}},
  \bibinfo{author}{\bibfnamefont{H.}~\bibnamefont{Tian}},
  \bibinfo{author}{\bibfnamefont{T.~R.} \bibnamefont{Shrout}},
  \bibinfo{author}{\bibfnamefont{Z.}~\bibnamefont{Xu}}, \bibnamefont{et~al.},
  \bibinfo{journal}{Nature} \textbf{\bibinfo{volume}{577}},
  \bibinfo{pages}{350} (\bibinfo{year}{2020}).

\bibitem[{\citenamefont{Pan et~al.}(2019)\citenamefont{Pan, Li, Liu, Zhang,
  Wang, Lan, Zheng, Ma, Gu, Shen et~al.}}]{pan2019ultrahigh}
\bibinfo{author}{\bibfnamefont{H.}~\bibnamefont{Pan}},
  \bibinfo{author}{\bibfnamefont{F.}~\bibnamefont{Li}},
  \bibinfo{author}{\bibfnamefont{Y.}~\bibnamefont{Liu}},
  \bibinfo{author}{\bibfnamefont{Q.}~\bibnamefont{Zhang}},
  \bibinfo{author}{\bibfnamefont{M.}~\bibnamefont{Wang}},
  \bibinfo{author}{\bibfnamefont{S.}~\bibnamefont{Lan}},
  \bibinfo{author}{\bibfnamefont{Y.}~\bibnamefont{Zheng}},
  \bibinfo{author}{\bibfnamefont{J.}~\bibnamefont{Ma}},
  \bibinfo{author}{\bibfnamefont{L.}~\bibnamefont{Gu}},
  \bibinfo{author}{\bibfnamefont{Y.}~\bibnamefont{Shen}}, \bibnamefont{et~al.},
  \bibinfo{journal}{Science} \textbf{\bibinfo{volume}{365}},
  \bibinfo{pages}{578} (\bibinfo{year}{2019}).

\bibitem[{\citenamefont{Pan et~al.}(2021)\citenamefont{Pan, Lan, Xu, Zhang,
  Yao, Liu, Meng, Guo, Gu, Yi et~al.}}]{pan2021ultrahigh}
\bibinfo{author}{\bibfnamefont{H.}~\bibnamefont{Pan}},
  \bibinfo{author}{\bibfnamefont{S.}~\bibnamefont{Lan}},
  \bibinfo{author}{\bibfnamefont{S.}~\bibnamefont{Xu}},
  \bibinfo{author}{\bibfnamefont{Q.}~\bibnamefont{Zhang}},
  \bibinfo{author}{\bibfnamefont{H.}~\bibnamefont{Yao}},
  \bibinfo{author}{\bibfnamefont{Y.}~\bibnamefont{Liu}},
  \bibinfo{author}{\bibfnamefont{F.}~\bibnamefont{Meng}},
  \bibinfo{author}{\bibfnamefont{E.-J.} \bibnamefont{Guo}},
  \bibinfo{author}{\bibfnamefont{L.}~\bibnamefont{Gu}},
  \bibinfo{author}{\bibfnamefont{D.}~\bibnamefont{Yi}}, \bibnamefont{et~al.},
  \bibinfo{journal}{Science} \textbf{\bibinfo{volume}{374}},
  \bibinfo{pages}{100} (\bibinfo{year}{2021}).

\bibitem[{\citenamefont{Disa et~al.}(2020)\citenamefont{Disa, Fechner, Nova,
  Liu, F{\"o}rst, Prabhakaran, Radaelli, and Cavalleri}}]{disa2020polarizing}
\bibinfo{author}{\bibfnamefont{A.~S.} \bibnamefont{Disa}},
  \bibinfo{author}{\bibfnamefont{M.}~\bibnamefont{Fechner}},
  \bibinfo{author}{\bibfnamefont{T.~F.} \bibnamefont{Nova}},
  \bibinfo{author}{\bibfnamefont{B.}~\bibnamefont{Liu}},
  \bibinfo{author}{\bibfnamefont{M.}~\bibnamefont{F{\"o}rst}},
  \bibinfo{author}{\bibfnamefont{D.}~\bibnamefont{Prabhakaran}},
  \bibinfo{author}{\bibfnamefont{P.~G.} \bibnamefont{Radaelli}},
  \bibnamefont{and}
  \bibinfo{author}{\bibfnamefont{A.}~\bibnamefont{Cavalleri}},
  \bibinfo{journal}{Nature Physics} \textbf{\bibinfo{volume}{16}},
  \bibinfo{pages}{937} (\bibinfo{year}{2020}).

\bibitem[{\citenamefont{Nova et~al.}(2019)\citenamefont{Nova, Disa, Fechner,
  and Cavalleri}}]{nova2019metastable}
\bibinfo{author}{\bibfnamefont{T.}~\bibnamefont{Nova}},
  \bibinfo{author}{\bibfnamefont{A.}~\bibnamefont{Disa}},
  \bibinfo{author}{\bibfnamefont{M.}~\bibnamefont{Fechner}}, \bibnamefont{and}
  \bibinfo{author}{\bibfnamefont{A.}~\bibnamefont{Cavalleri}},
  \bibinfo{journal}{Science} \textbf{\bibinfo{volume}{364}},
  \bibinfo{pages}{1075} (\bibinfo{year}{2019}).

\bibitem[{\citenamefont{Budden et~al.}(2021)\citenamefont{Budden, Gebert,
  Buzzi, Jotzu, Wang, Matsuyama, Meier, Laplace, Pontiroli, Ricc{\`o}
  et~al.}}]{budden2021evidence}
\bibinfo{author}{\bibfnamefont{M.}~\bibnamefont{Budden}},
  \bibinfo{author}{\bibfnamefont{T.}~\bibnamefont{Gebert}},
  \bibinfo{author}{\bibfnamefont{M.}~\bibnamefont{Buzzi}},
  \bibinfo{author}{\bibfnamefont{G.}~\bibnamefont{Jotzu}},
  \bibinfo{author}{\bibfnamefont{E.}~\bibnamefont{Wang}},
  \bibinfo{author}{\bibfnamefont{T.}~\bibnamefont{Matsuyama}},
  \bibinfo{author}{\bibfnamefont{G.}~\bibnamefont{Meier}},
  \bibinfo{author}{\bibfnamefont{Y.}~\bibnamefont{Laplace}},
  \bibinfo{author}{\bibfnamefont{D.}~\bibnamefont{Pontiroli}},
  \bibinfo{author}{\bibfnamefont{M.}~\bibnamefont{Ricc{\`o}}},
  \bibnamefont{et~al.}, \bibinfo{journal}{Nature Physics}
  \textbf{\bibinfo{volume}{17}}, \bibinfo{pages}{611} (\bibinfo{year}{2021}).

\end{thebibliography}
\end{document}